\documentclass{aa}  
\usepackage{lineno}
\renewcommand{\linenumbers}{}
\usepackage{natbib}
\bibpunct{(}{)}{;}{a}{}{,} 
\usepackage{graphicx}
\usepackage{txfonts}
\usepackage{placeins}
\usepackage{lipsum}
\usepackage{float}
\usepackage{subcaption}        
\usepackage{lscape}
\usepackage{placeins}
\usepackage{hyperref}

\begin{document}

   \title{A \textit{JWST} project on 47\,Tucanae: kinematics, energy equipartition and anisotropy of multiple populations}

   \author{T. Ziliotto \inst{1}
        \and A. P. Milone\inst{1,2}
        \and G. Cordoni\inst{3}
        \and F. I. Aros\inst{4}
        \and E. Vesperini\inst{5}
        \and J.-W. Lee\inst{6}
        \and A. Bellini\inst{7}
        \and P. Bianchini\inst{8}
        \and A. Mastrobuono-Battisti\inst{1}
        \and M. Libralato\inst{2}
        \and E. Dondoglio\inst{2}
        \and M. Tailo\inst{1}
        \and A. Livernois\inst{5}
        \and M. V. Legnardi\inst{1}
        \and E. Lagioia\inst{9}
        \and E. Bortolan\inst{1}
        \and F. Muratore\inst{1}
        \and A. F. Marino\inst{2}
        \and A. Alves-Brito\inst{10}
        \and A. Renzini\inst{2}
        }

   \institute{Dipartimento di Fisica e Astronomia Galileo Galilei, Univ. di Padova, Vicolo dell'Osservatorio 3, Padova, IT-35122\\
             \email{tuila.ziliotto@unipd.it}
    \and Istituto Nazionale di Astrofisica - Osservatorio Astronomico di Padova, Vicolo dell'Osservatorio 5, Padova, IT-35122
    \and Research School of Astronomy \& Astrophysics, Australian National University, Canberra, ACT 2611, Australia
    \and Department of Astrophysics, University of Vienna, Türkenschanzstrasse 17, 1180 Vienna, Austria
    \and Department of Astronomy, Indiana University, Bloomington, IN 47401, USA
    \and Department of Physics and Astronomy, Sejong University, 209 Neungdong-ro, Gwangjin-Gu, Seoul, 05006, Republic of Korea
    \and Space Telescope Science Institute, 3700 San Martin Dr, Baltimore, MD, 21218, USA
    \and Université de Strasbourg, CNRS, Observatoire astronomique de Strasbourg, UMR 7550, F-67000 Strasbourg, France
    \and South-Western Institute for Astronomy Research, Yunnan University, Kunming 650500, P. R. China
    \and  Universidade Federal do Rio Grande do Sul, Instituto de Física, Av. Bento Gonçalves 9500, Porto Alegre, RS, Brazil
    }

  \abstract{Recent work with \textit{JWST} has demonstrated its capability to identify and chemically characterize multiple populations in globular clusters down to the H-burning limit. In this study, we explore the kinematics of multiple populations in the globular cluster 47\,Tucanae by combining data from \textit{JWST}, \textit{HST}, Gaia, and ground-based telescopes. 
  We analyzed velocity dispersion and anisotropy profiles from the cluster center out to $\sim$10$R_h$. Our findings indicate that while first population (1G) stars' motions are isotropic, second population (2G) stars' motions are significantly radially anisotropic. 
  These results align with the predictions of simulations of the dynamical evolution of clusters where 2G stars are initially more centrally concentrated than 1G stars.
  Furthermore, we subdivided the 2G population into two subpopulations: $2G_A$ and $2G_B$, with the latter being more chemically extreme. We compared their dynamical profiles and found no significant differences.
  For the first time, we measured the degree of energy equipartition among the multiple populations of 47\,Tucanae. Overall, within the analyzed radial range ($\sim$2-4$R_h$), both populations exhibit a low degree of energy equipartition. 
  The most significant differences between 1G and 2G stars are observed in the tangential velocity component, where 2G stars are characterized by a stronger degree of energy equipartition than 1G stars.
  In the radial component, the behavior of 1G and 2G stars is more variable, with differences largely dependent on radius.
  Moreover, our analysis reveals that the ratio of rotational velocity to velocity dispersion is larger for the 2G population. Finally, we found that 1G stars exhibit higher skewness in their tangential proper motions than 2G stars, providing additional evidence of kinematic differences between the two stellar generations.

  }


   \maketitle

\section{Introduction} \label{sec:intro}

Nearly all Galactic globular clusters (GCs) exhibit star-to-star variations in their light element abundances. While a significant fraction of the stars presents a chemical composition similar to $\alpha$-enhanced halo field stars with the same metallicity (first population, 1G), the remaining stars (second population, 2G) display depletion in C and O and enhancement in He, Na, N, and Al \citep[see][for recent reviews]{bastian2018a,gratton2019a,milone2022a}.

The origin of this phenomenon has been widely discussed, and multiple scenarios have been proposed to explain the presence of multiple populations \citep{cottrell1981a,dantona1983a,decressin2007a,bastian2013a,renzini2015a,dantona2016a,dercole2016,calura2019a,wang2020a, renzini2022a}. Recent studies using data from the James Webb Space Telescope (\textit{JWST}) and the Hubble Space Telescope (\textit{HST}) have analyzed the chemical variations between 1G and 2G very low-mass stars in M\,92 and 47\,Tucanae \citep{ziliotto2023, milone2023a, marino2024a, marino2024b} and appear to support models predicting multiple episodes of star formation within the same cluster.

A number of studies based on hydro/N-body simulations have investigated the formation of multiple generations in globular clusters and predicted that 2G stars form centrally concentrated in the inner regions of the 1G system \citep[see e.g.][]{dercole2008,bekki2011,bekki2017,calura2019a,lacchin2021,lacchin2022,yaghoobi2022a,yaghoobi2022b}.
While long-term dynamical evolution can gradually erase information of the initial conditions of these systems, some clusters can still preserve this information \citep[see e.g.][]{vesperini2021,livernois2024}, as numerous observational studies demonstrate \citep{richer2013,bellini2015a,bellini2018,milone2018a,dalessandro2019,dalessandro2024,libralato2023a,leitinger2023,cordoni2020a,cordoni2025,mehta2025}.

As the cluster evolves due to the effects of two-body relaxation, stellar interactions and energy exchange drive the system toward energy equipartition, a dynamical state where at a given distance from the cluster center stars with different masses have equal kinetic energy, $K=\frac{1}{2}m\sigma^2$, and thus a velocity dispersion, $\sigma$, increasing for smaller stellar masses \citep[see e.g.][]{spitzer1969}.
Theoretical and observational works suggest that GCs do not reach full energy equipartition, but are characterized by a state of partial equipartition \citep[see e.g.][]{trenti2013,bianchini2016,libralato2018,watkins2022,pavlik2021,pavlik2022,aros2023}.

The structural and kinematic properties of multiple populations contain key information on their formation and dynamical history. In two recent numerical studies of the dynamics of multiple populations, \cite{vesperini2021} and \cite{livernois2024}  have shown that the initial differences between the spatial concentration of 1G and 2G stars lead to differences in their subsequent evolution towards energy equipartition resulting in differences between the present-day degree of equipartition of the 1G and the 2G population.

However, measuring the degree of energy equipartition of the multiple populations in GCs is challenging. Generally, high-precision proper motion measurements in GCs are more commonly available for bright stars, such as those on the red giant branch (RGB), asymptotic giant branch (AGB), and horizontal branch (HB), which limits the analysis to a narrow mass range. This excludes the entire main sequence (MS), which encompasses a significantly broader range of stellar masses compared to giant stars. Furthermore, high-precision photometry with filters that effectively separate multiple populations is also required. 

In the past few years, thanks to progress in measuring kinematics of low-mass stars with the \textit{HST}, the measurement of energy equipartition in GCs has become possible. \cite{watkins2022} measured the degree of energy equipartition in the central regions of nine GCs by using photometric and astrometric data from the Hubble Space Telescope Proper Motion (\textit{HSTPROMO}) catalogs of Galactic GCs, finding that no cluster is at or near full equipartition. Similarly, \cite{libralato2022a} found a partial state of energy equipartition in NGC\,5904. 
\cite{milone2025} explored the degree of energy equipartition in NGC 6397, finding that stars in the cluster exhibit only a very small degree of energy equipartition.
\cite{heyl2017} investigated mass-dependent kinematics in 47\,Tucanae, finding a weak dependence of proper motions on stellar mass just beyond the cluster's half-light radius, and a stronger dependence at the cluster center.
In the context of multiple stellar populations, only one cluster has been studied, and within a small radial range. \cite{bellini2018} measured the degree of energy equipartition for 1G and 2G stars in $\omega$ Centauri, finding that 2G stars present some degree of energy equipartition, while 1G stars are not in equipartition. 

Recently, we initiated a project to investigate the GC 47\,Tucanae and its stellar populations using \textit{JWST} observations from GO 2560 (PI: A. F. Marino) and archival data from major telescopes, including \textit{HST}, \textit{JWST}, and Gaia. Early results from this project include:
   i) the identification and chemical characterization of multiple stellar populations among very-low-mass stars using deep Near Infrared Camera (NIRCam) photometry \citep{milone2023a, marino2024a, legnardi2024a}; 
    ii) the first detection of the brown dwarf sequence and the discovery of gaps and discontinuities within it \citep{marino2024a};
    iii) the inaugural spectroscopic study of multiple populations among M-dwarfs, based on Near-Infrared Spectrograph (NIRSpec) data \citep{marino2024b}; and
    iv) the exploration of binary systems among first-generation and second-generation stars \citep{milone2025}.

In this paper, we continue our investigation into the multiple stellar populations of 47\,Tucanae
 focusing on the kinematics of these populations. By leveraging a combination of datasets from \textit{JWST}, \textit{HST}, Gaia, and ground-based telescopes, we identify stellar populations across a mass range from $\sim$0.1 to 0.9 $M_{\odot}$ and derive precise proper motions for these stars. For the first time, we measure the differences in the degree of energy equipartition between 1G and 2G stars in 47\,Tucanae, within an extensive radial range from approximately 1.5 to 4.5 $R_h$. We further analyze the internal kinematics of 1G and 2G stars, including their anisotropy profiles, over an extensive radial range that extends from the center of the cluster to over 30 arcminutes. This paper is organized as follows: Section \ref{sec:data} describes our photometric and astrometric catalogs; in Section \ref{sec:mass}, we outline our procedure for deriving stellar masses; the methodology and results regarding velocity dispersion and anisotropy profiles are discussed in Section \ref{sec:ani}; Section \ref{sec:equi} presents the energy equipartition models and results; and Section \ref{sec:sd} provides a discussion of the results, including energy equipartition radial profiles for 1G and 2G stars.

\section{Data} \label{sec:data}

\begin{figure}
    \centering
    \includegraphics[width=1\linewidth]{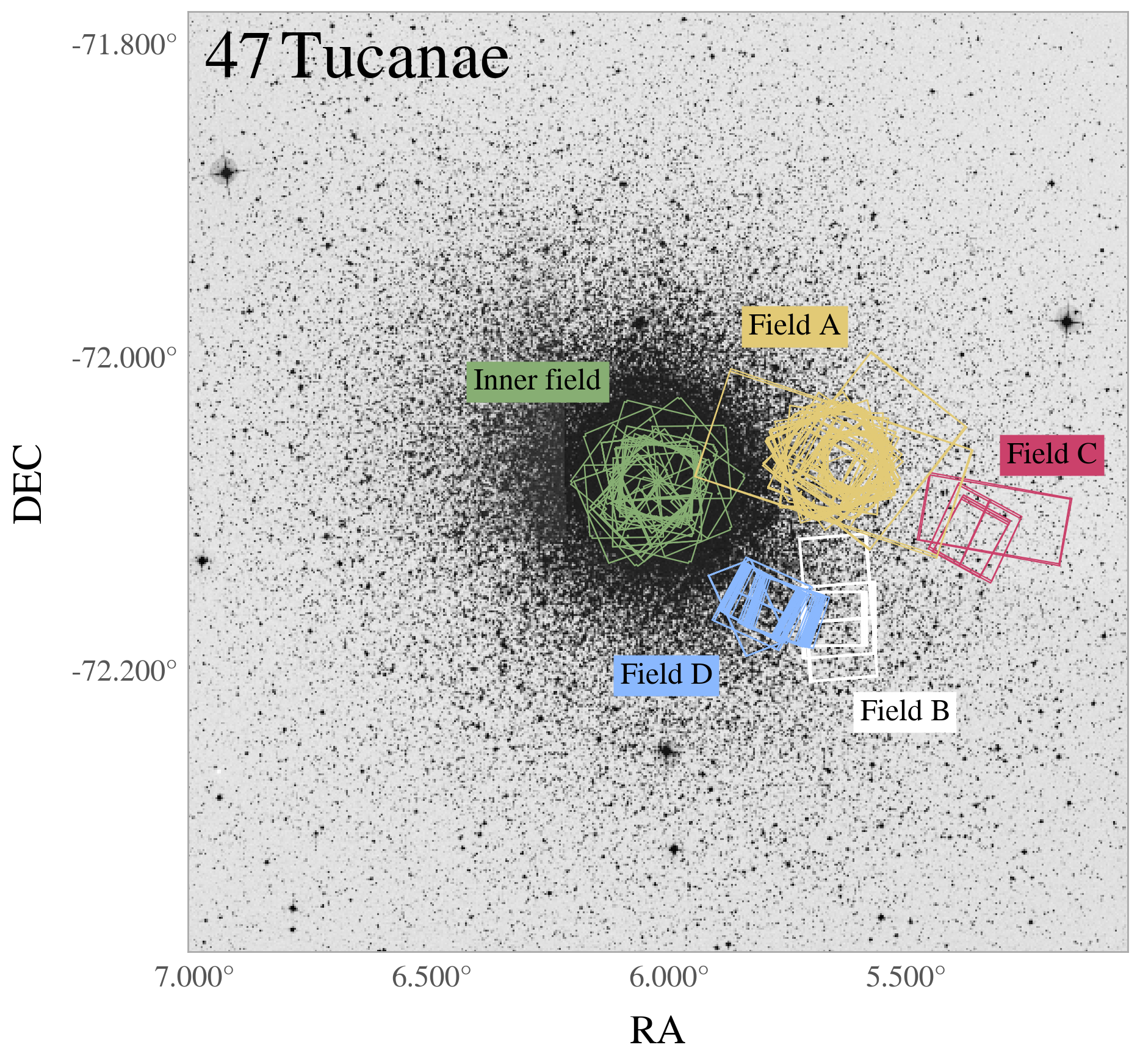}
    \caption{Footprints of the \textit{JWST} (fields A, B, and C) and \textit{HST} (fields A, B, C, D, and inner field) observations, with north at the top and east to the left.}
    \label{fig:footprint}
\end{figure}

\begin{figure}
    \centering
    \includegraphics[width=1\linewidth]{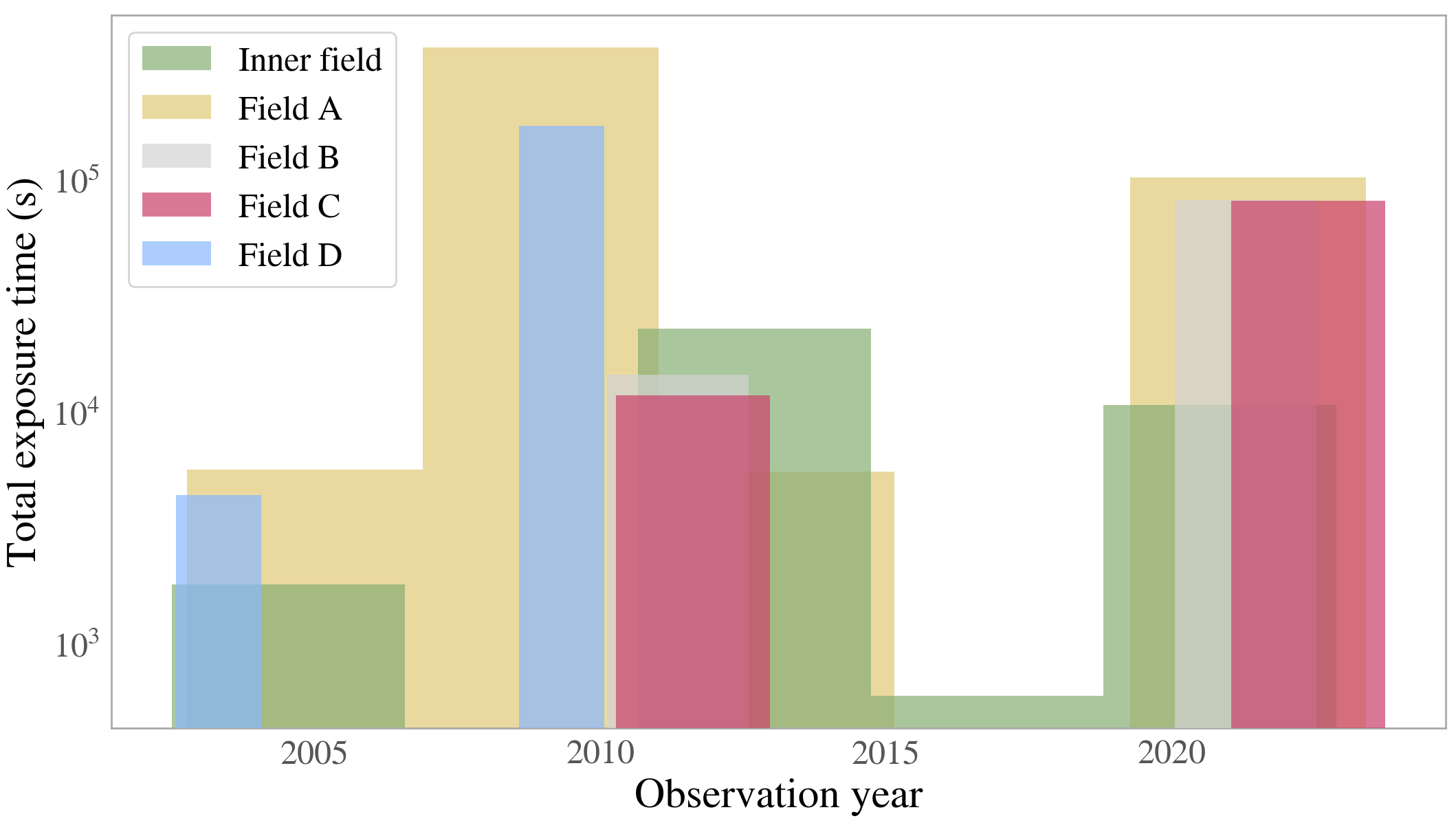}
    \caption{Total exposure time per observation year for fields A, B, C, D, and the inner field.}
    \label{fig:exptime}
\end{figure}

In this section, we describe the dataset used to identify multiple stellar populations in the various fields of view analyzed in this study, illustrated in Figure \ref{fig:footprint}. The cumulative exposure times across different years of observation are illustrated in the histogram in Figure \ref{fig:exptime}. The data for fields A, B, C, D, and the inner field are available in the Mikulski Archive for Space Telescopes (MAST)\footnote{Inner field: \href{http://dx.doi.org/10.17909/gn0t-6945}{http://dx.doi.org/10.17909/gn0t-6945};\\ Field A: \href{http://dx.doi.org/10.17909/vha2-8275}{http://dx.doi.org/10.17909/vha2-8275};\\ Field B: \href{http://dx.doi.org/10.17909/dyjj-z730}{http://dx.doi.org/10.17909/dyjj-z730};\\ Field C: \href{http://dx.doi.org/10.17909/bcf5-c348}{http://dx.doi.org/10.17909/bcf5-c348};\\ Field D: \href{http://dx.doi.org/10.17909/0abv-jg85}{http://dx.doi.org/10.17909/0abv-jg85}.}.

For the inner field, we used the astro-photometric catalog from \cite{libralato2022a,libralato2023a} to analyze the internal kinematics of multiple stellar populations. This catalog employs \textit{HST} UV and optical photometry to distinguish between different populations and derive proper motions. Moreover, we combined the astrometric data from this catalog with the multiple population catalog of HB stars from \cite{dondoglio2021a} in order to have proper motions for 1G and 2G HB stars in the central region of 47\,Tucanae.  We refer to these works for a detailed description of their methodology.

Field A, situated approximately 7 arcminutes from the cluster center, has been extensively observed with the \textit{HST}, providing deep optical and near-infrared data. With this dataset, we were able to identify multiple populations among lower- ($M<0.45M_\odot$) and upper-MS ($0.67 < M/M_{\odot}<0.78$) stars.
This dataset was previously used by \cite{milone2023a}, which presents a detailed study of the chemical composition of multiple populations among very low-mass stars in 47\,Tucanae.

Fields B and C, situated at $\sim 8.5$ and $\sim 11$ arcminutes from the cluster center, respectively, were observed with \textit{JWST}/NIRCam as part of the program GO-2560 (PI: A. F. Marino). To classify populations in these fields, we combined NIRCam photometry in the F115W and F322W2 filters with observations taken with the \textit{HST} Ultraviolet-VISible (UVIS) channel of the Wide Field Camera 3 (WFC3) in the F606W filter, and with NIR/WFC3 in the F110W and F160W filters. See \cite{marino2024a} and \cite{milone2023a} for a detailed description of these datasets and a study of multiple populations among very low-mass stars and brown dwarfs in 47\,Tucanae.

Field D, situated at approximately 6 arcminutes from the cluster center, comprises deep NIR/WFC3 \textit{HST} observations in the F110W and F160W filters, UVIS/WFC3 observations in the F606W and F390W filters, and Wide Field Channel (WFC) observations taken with the Advanced Camera for Surveys (ACS) in the F775W filter.
Similar to \textit{JWST} filters, these filters are capable of disentangling multiple populations because they are sensitive to molecules containing oxygen. Since the 2G stars are oxygen depleted, they exhibit brighter F160W magnitudes and redder F110W $-$ F160W colors compared to the 1G stars \citep[see e.g.][]{dondoglio2022a}.

To measure stellar fluxes and positions in HST images, we followed a two-step procedure. First-step photometry for infrared images was performed using the FORTRAN code \textit{hst1pass} \citep{anderson2022a} to find and measure bright and unsaturated sources in the individual exposures. For optical images, we generally used the \textit{img2xym} computer program. Originally developed by Jay Anderson for HST image reduction \citep[e.g.,][]{anderson2006a}, this approach measures stellar fluxes and positions in each image separately, using a spatially variable PSF model \citep{anderson2000a} along with a perturbation PSF to account for subtle focus variations. The perturbation PSF was derived from bright, unsaturated, and isolated stars. Photometric measurements from different exposures and detectors were standardized to a common zero-point by referencing the deepest exposure in each filter. This was achieved by using well-measured bright stars to compute the magnitude offsets between individual exposures and the reference frame, allowing all measurements to be transformed into a common master frame. Geometric distortion corrections were applied using the solutions from \cite{milone2023a}. Stellar positions were transformed into a common reference frame aligned with the Gaia DR3 catalog, ensuring that the x- and y-coordinates corresponded to west and north, respectively.

Second-step photometry was performed using the FORTRAN-based software KS2, developed by Jay Anderson as an advancement of the kitchen sinc algorithm \citep{anderson2008a}. KS2 implements three distinct techniques for stellar measurements. Method I, optimal for bright stars, determines magnitudes and positions by fitting the most suitable effective Point Spread Function (PSF) model \citep[e.g.,][]{anderson2000a}. These values are computed individually for each image and subsequently averaged to refine the final measurements. Method II, designed to optimize astrometry and photometry for faint sources, integrates information from multiple exposures simultaneously. In this approach, fluxes are derived by subtracting neighboring stars and performing aperture photometry within a 5 $\times$ 5 pixel grid. Method III operates similarly to Method II but employs circular aperture photometry with a 0.75-pixel radius, making it particularly effective in crowded fields \citep[see][for details]{sabbi2016a,bellini2017a,milone2022b}. Photometric calibration follows the procedures in \cite{milone2022b} and utilizes the photometric zero-points available on the Space Telescope Science Institute (STScI) webpage for WFC/ACS, UVIS/WFC3, and NIR/WFC3\footnote{\href{https://www.stsci.edu/hst/instrumentation/acs/data-analysis/zeropoints}{https://www.stsci.edu/hst/instrumentation/acs/data-analysis/zeropoints}, \href{https://www.stsci.edu/hst/instrumentation/wfc3/data-analysis/photometric-calibration}{https://www.stsci.edu/hst/instrumentation/wfc3/data-analysis/photometric-calibration}}. Geometric distortion corrections are applied using the solutions from \cite{anderson2006a} and \cite{bellini2011a} for WFC/ACS and UVIS/WFC3, while NIR/WFC3 corrections follow \cite{anderson2022a}. To ensure high-quality photometry and astrometry, we implemented the selection criteria outlined by \cite[][see their Section 2.4]{milone2022b}, leveraging KS2 diagnostic parameters to identify well-isolated stars with minimal root mean scatter in position and magnitude.

Similarly, for JWST/NIRCam data reduction, we performed the same two-step procedure. First-step photometry was obtained using the \textit{img2xym} computer program described above. 
Second-step photometry was performed using KS2 \citep[see][for details]{milone2023a,marino2024a}. 
The final photometry was calibrated to the VEGA system using the zero-points provided on the STScI webpage for NIRCam\footnote{\href{https://jwst-docs.stsci.edu/jwst-near-infrared-camera/nircam-performance/nircam-absolute-flux-calibration-and-zeropoints}{https://jwst-docs.stsci.edu/jwst-near-infrared-camera/nircam-performance/nircam-absolute-flux-calibration-and-zeropoints}}.

To analyze the kinematics of more massive 1G and 2G stars, we used three catalogs: \cite{cordoni2025}, \cite{mehta2025}, and \cite{lee2022a}. \cite{mehta2025} used Gaia DR3 XP spectra \citep{gaia2023b} to derive synthetic photometric filters that distinguish multiple populations in the RGB, HB, and AGB out to the tidal radius of 47\,Tucanae. Using ground-based photometry from \cite{stetson2019}, \cite{cordoni2025} applied the $\Delta_{C_{U,B,I}}$ vs. $\Delta_{B,I}$ chromosome map (ChM) to separate 1G and 2G RGB stars in 47\,Tucanae, out to distances of about 20 arcminutes. Finally, \cite{lee2022a} compiled large field-of-view ($\sim 1^{\circ} \times 1^{\circ}$) ground-based photometry in filters designed to maximize the separation of 1G and 2G stars. We matched this catalogs with Gaia DR3 proper motions to examine kinematic differences between 1G and 2G stars. Together, the catalogs from \citet{lee2022a, mehta2025, cordoni2025} cover stars from about 5 arcminutes to the tidal radius of 47\,Tucanae.

\begin{figure*}
    \centering
    \includegraphics[width=1\textwidth]{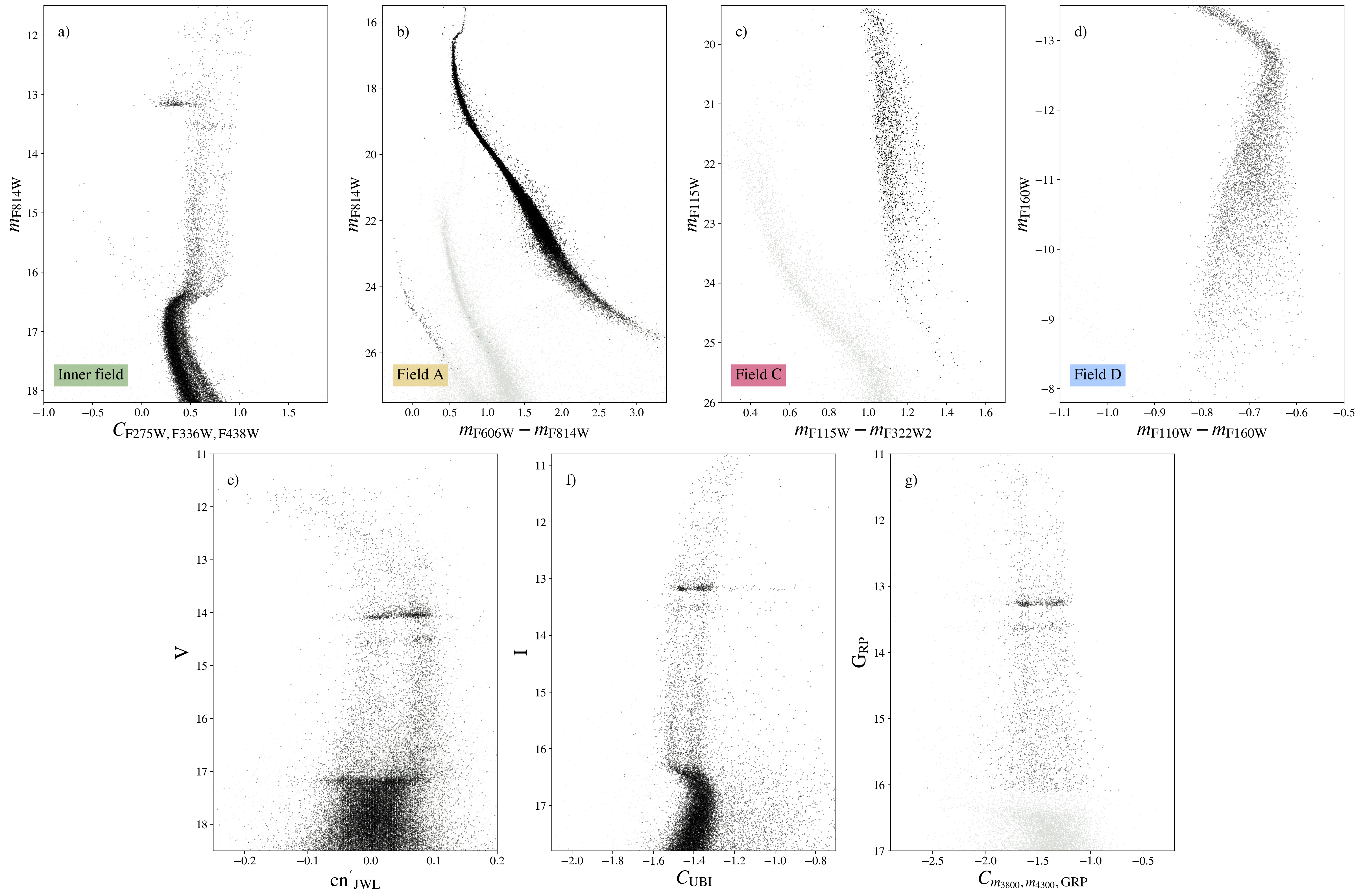}

     \caption{Photometric diagrams demonstrating the photometry used to disentangle the multiple populations in different radial regions of 47\,Tucanae. Panel a) shows the \textit{HST} UV photometry used to identify populations in the RGB. In panel b), we show the CMD created using \textit{HST} optical filters for Field A. Panel c) demonstrates the CMD created for the lower MS and substellar objects with \textit{JWST} filters for Field C. Panel d) displays \textit{HST} IR photometry for disentangling populations among the lower MS. In panel e) we show Ca-CN photometry by \cite{lee2022a}, which is effective in identifying populations among RGB and HB stars. Panel f) displays the diagram with ground-based UBVI photometry from \cite{stetson2019}. Finally, in panel g), we show the CMD created with Gaia XP spectro-photometry \citep{mehta2025}.}
    \label{fig:cmds}
\end{figure*}

\begin{figure*}
    \centering

    \includegraphics[width=1\textwidth]{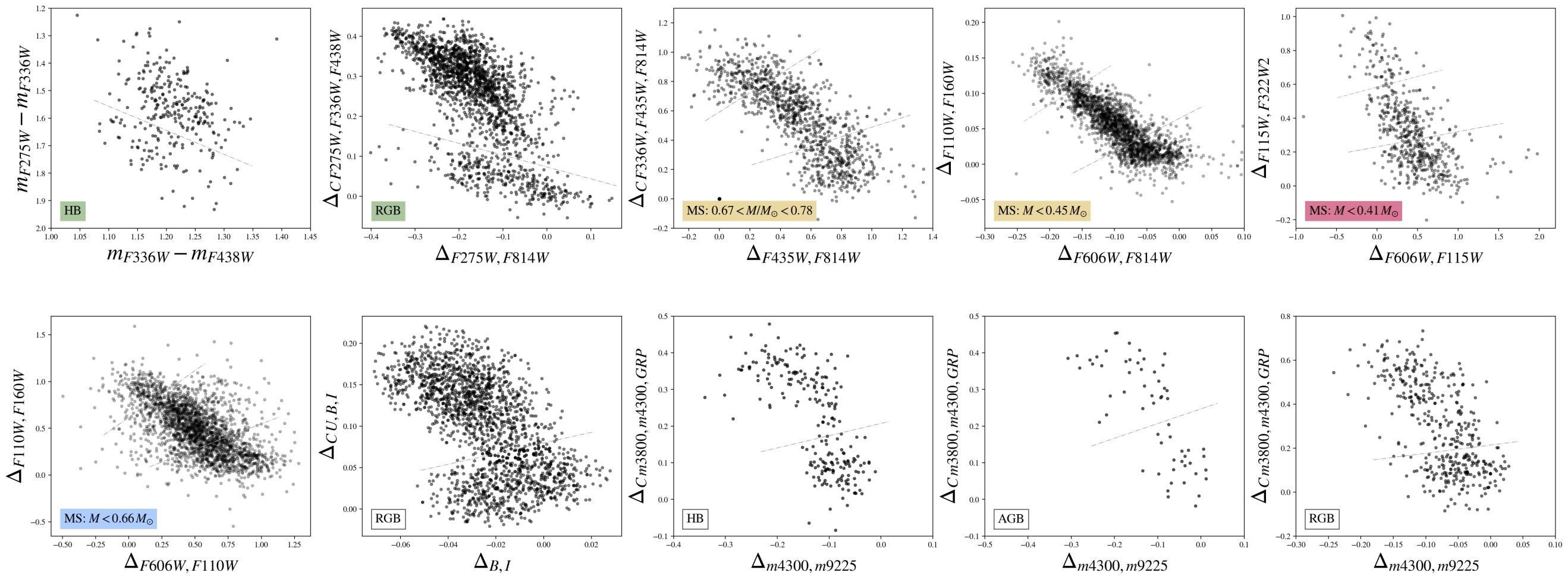}
    
    \caption{Collection of ChMs of stars in different evolutionary stages in 47\,Tucanae (RGB, AGB, HB, upper MS, and lower MS). The labels are color-coded according to the different observed fields, following the same color scheme as Figure \ref{fig:cmds}.} 
    \label{fig:chms}
\end{figure*}

Figure \ref{fig:cmds} displays the color-magnitude diagrams (CMDs) created from the photometric data of the previously discussed fields of view. The presence of distinct stellar populations in 47 Tucanae is clearly evident in the CMDs, which reveal multiple stellar sequences. Similarly, the ChMs presented in Figure \ref{fig:chms} further highlight the existence of these multiple populations for stars at various evolutionary stages.

\section{Proper motions} \label{subsec:propermotions}

To assess relative proper motions for \textit{HST} and \textit{JWST} data, we employed a methodology detailed in several previous studies \citep[e.g.,][]{anderson2003a, piotto2012a, libralato2022a, milone2023a}, which focuses on comparing stellar positions across images captured at different epochs.

\begin{figure}
    \centering
    \includegraphics[width=\linewidth]{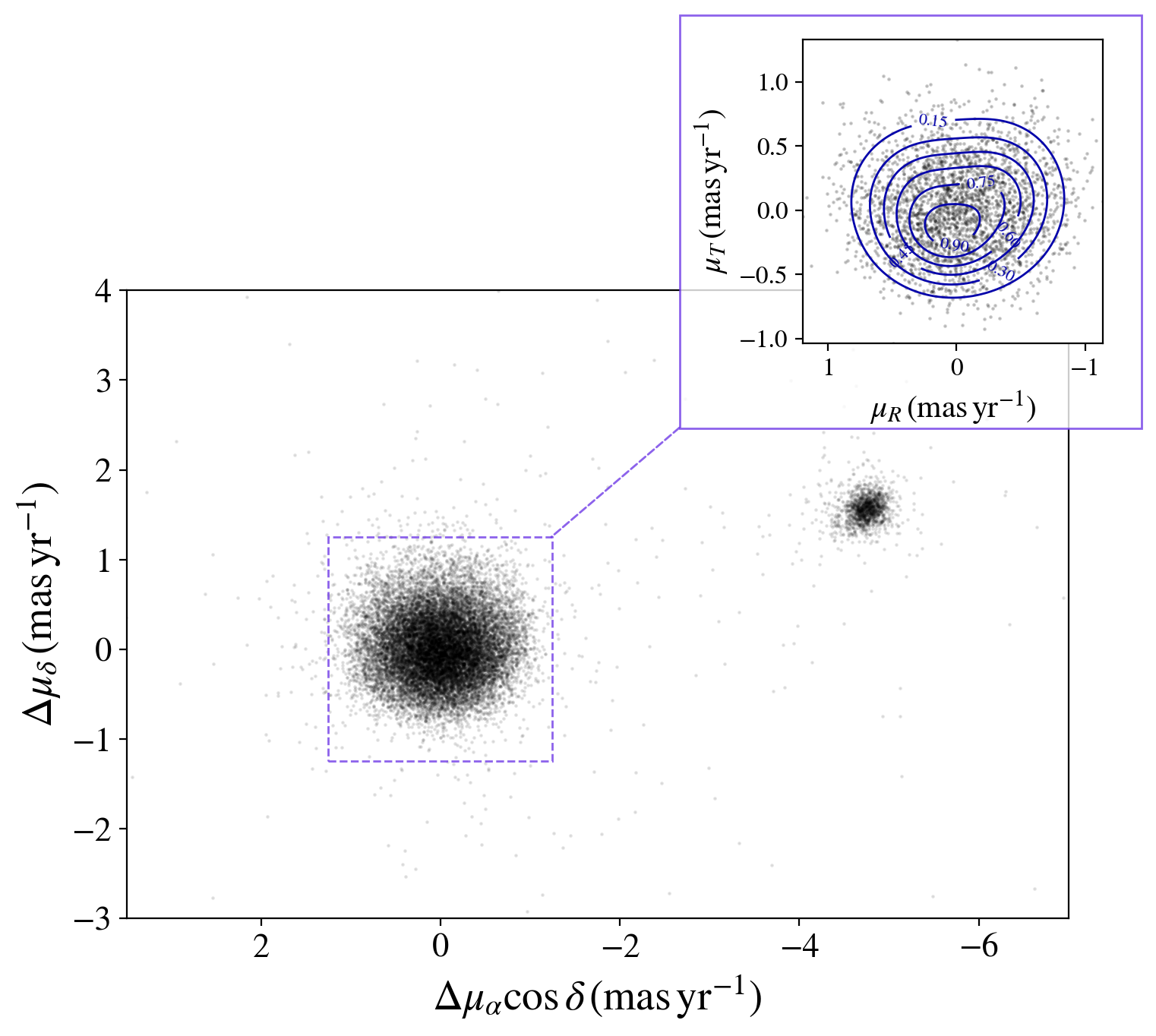}
    \caption{Proper motions of stars in Field A, derived using multi-epoch \textit{HST} observations. The small clustering of stars in the top left region corresponds to the Small Magellanic Cloud. The subplot on the top right displays the projected proper motions in tangential and radial directions. The blue contour lines represent the probability density estimated by a Gaussian Mixture Model, showcasing the statistical distribution and asymmetries in the data.}
    \label{fig:pms}
\end{figure}

Our first step involved creating distinct photometric and astrometric catalogs from images taken in multiple filters over several epochs. The master frame for our study was established using the first-epoch images obtained with the reddest filter. We oriented this reference frame with the X-axis directed west and the Y-axis directed north.

To transform star coordinates from each catalog into this master frame, we utilized six-parameter linear transformations.
The transformation process relied on bright, unsaturated cluster stars. This selection was conducted in two phases: initially, we identified likely members of 47\,Tucanae based solely on their positions in the CMD, which allowed us to derive preliminary proper motions. In the second phase, we refined our criteria by examining each star's position within both the CMD and the proper motion diagram, leading to a more accurate identification of cluster members. These members were then used to calculate improved proper motions. To further reduce uncertainties in our measurements, we adjusted stellar coordinates at each epoch by accounting for displacements due to proper motions. These corrected coordinates were then utilized to calculate improved transformations, further refining our estimates of proper motion.

To enhance alignment and minimize small residual distortions, accounting for CTE and geometric distortion residuals in the \textit{HST} (ACS and UVIS) images, we implemented local transformations based on the nearest N reference stars, deliberately excluding target stars themselves from influencing these corrections \citep[as described in][]{bellini2014}. The selection of N and the corresponding magnitude range was optimized to balance capturing local variations while maintaining a statistically significant sample of reference stars.
For stars near the extremes of the magnitude range, either very bright or very faint targets, we adjusted the magnitude selection criteria by using reference stars within the brightest or faintest magnitude bins instead of applying a fixed magnitude difference. This adaptive approach yielded a larger statistical sample and improved the correction process in regions where suitable reference stars were otherwise scarce.

We then plotted the x and y coordinates of each star against the observation epochs, fitting a weighted least-squares linear regression to these data points. The slope of this fitted line, along with its uncertainty, yields our best estimate of the star's proper motion.

As an example, Figure \ref{fig:pms} shows the proper motions for stars in Field A, derived by leveraging multi-epoch \textit{HST} observations.

\section{Stellar masses} \label{sec:mass}

\begin{figure}
    \centering
    \includegraphics[width=0.75\linewidth]{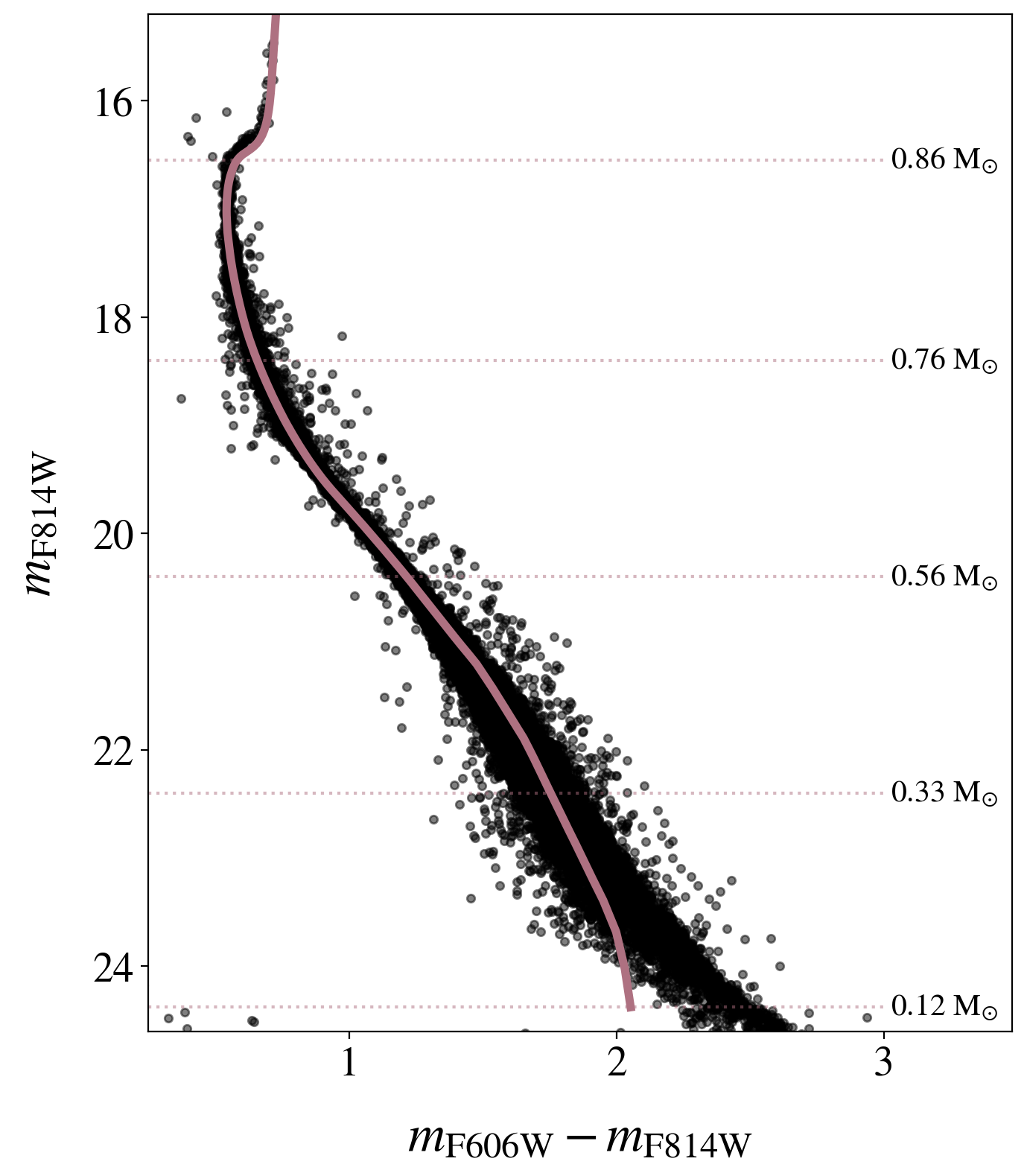}
    \caption{$m_{\rm F814W}$ vs. $m_{\rm F606W} - m_{\rm F814W}$ CMD of stars in Field A. The pink line represents the isochrone used to fit the cluster (see text for details), and the horizontal dotted lines indicate initial MS mass values along the CMD inferred from the isochrone.}
    \label{fig:mass}
\end{figure}

To analyze the mass dependence of the velocity dispersion, we estimated stellar masses for our various datasets using isochrone fitting. We employed isochrones from the Dartmouth Stellar Evolution Database \citep{dotter2008a} with the following parameters: $\mathrm{[Fe/H]} = -0.75$ dex, age $=$ 13 Gyr, helium mass fraction Y = 0.246, $\alpha$ enhancement [$\alpha$/Fe] $= +0.4$ dex, foreground reddening $E(B-V) = 0.03$ mag, and distance modulus $(m-M)_0 = 13.38$ mag, as in \cite{milone2023a}. As an example, Figure \ref{fig:mass} shows the CMD for Field A, constructed from \textit{HST} photometry. Stars are color-coded by their estimated mass, and the fitted isochrone is overlaid in black.

We utilized different CMDs to fit isochrones across our observational fields due to variations in available filter data. For Field A, we employed \textit{HST} photometry using $m_\mathrm{F814W}$ vs. $m_\mathrm{F606W}-m_\mathrm{F814W}$. Fields B and C were analyzed using \textit{JWST} photometry, specifically $m_\mathrm{F322W2}$ vs. $m_\mathrm{F115W}-m_\mathrm{F322W2}$. For Field D, we used \textit{HST} photometry with $m_\mathrm{F775W}$ vs. $m_\mathrm{F606W}-m_\mathrm{F775W}$. The Stetson photometry was analyzed using the $V$ vs. $V-I$ CMD. We excluded \cite{lee2022a} filters from our energy equipartition analysis due to the lack of available isochrones for these filters.

HB stars in 47\,Tucanae have masses of approximately 0.65 $M_{\odot}$ \citep{tailo2020}. Given that stars do not undergo significant mass loss during their HB evolution, AGB stars are assumed to have approximately the same mass. However, the mass loss that RGB stars experience before reaching the HB occurs on timescales much shorter than the two-body relaxation time. Consequently, AGB and HB stars are expected to retain kinematic properties characteristic of higher-mass stars. Thus, we assigned these stars the maximum mass measured in our RGB sample (as in \citeauthor{watkins2022} \citeyear{watkins2022}).

\section{Internal kinematics} \label{sec:ani}

To analyze the internal dynamics of the multiple populations in 47\,Tucanae, we combined our proper motion dataset (Section \ref{subsec:propermotions}) with the population classification described in Section \ref{sec:data}. Following the methodology of \cite{cordoni2025}, we converted stellar coordinates and proper motion components ($\alpha$, $\delta$, $\mu_{\alpha}\cos\delta$, $\mu_{\delta}$) to Cartesian coordinates (x, y, $\mu_x$, $\mu_y$) using Equation 2 from \cite{gaia2018}. To analyze stellar motions relative to the cluster for the Gaia dataset, we subtracted the cluster's motion as determined by \cite{vasiliev2021}. Subsequently, we projected the proper motions into radial ($\mu_\mathrm{R}$) and tangential ($\mu_\mathrm{T}$) components with respect to the cluster center.
We performed a correction in absolute $\mu_\mathrm{R}$ proper motions to account for perspective contraction/expansion due to the bulk motion of 47\,Tucanae along the line of sight \citep[Equation 4 of][]{bianchini2018b}.

\subsection{Rotation}
\label{sec:rot}
\begin{figure}
    \centering
    \includegraphics[width=0.9\linewidth]{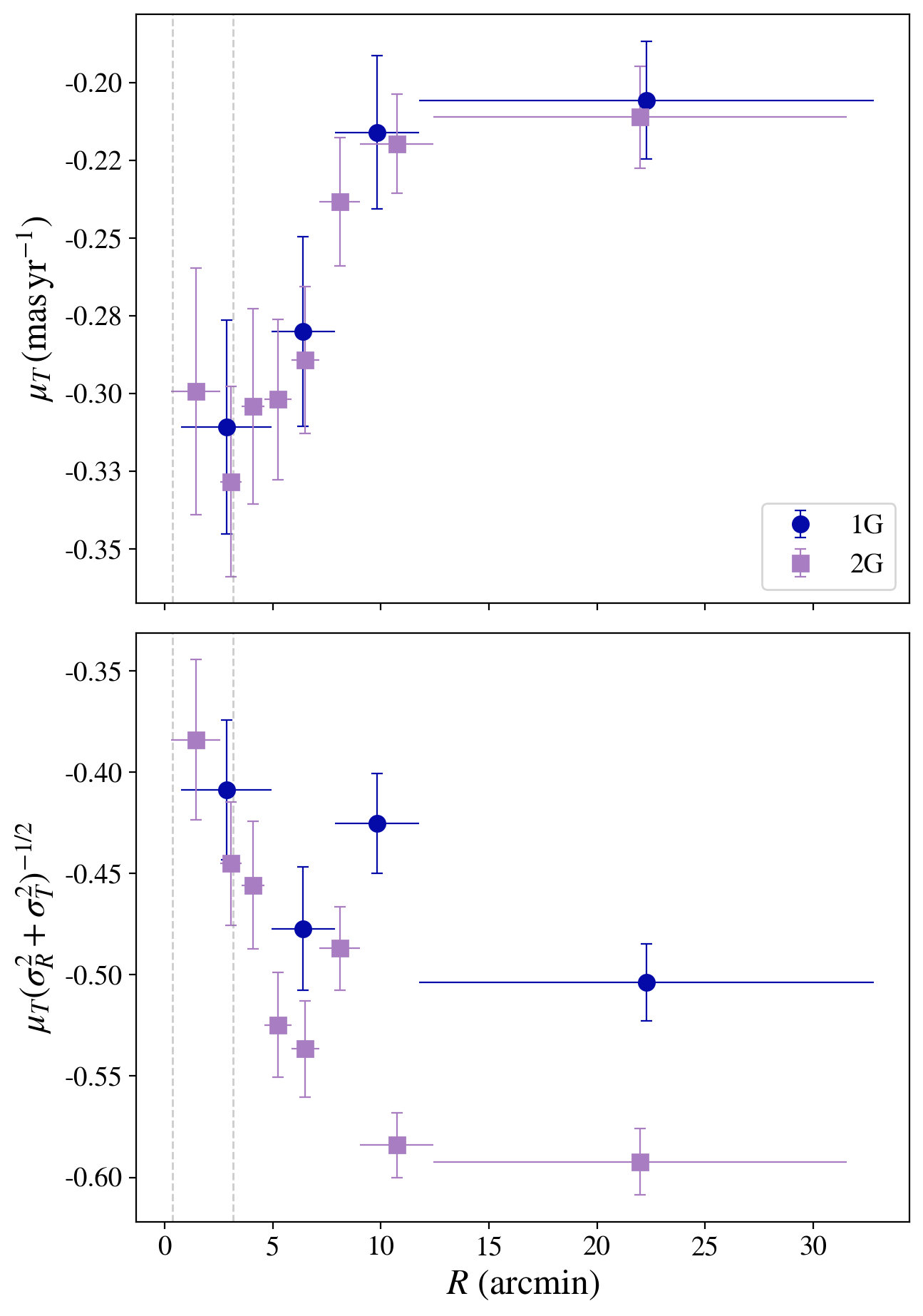}
    \caption{Top: Radial profiles of mean tangential velocity for 1G (blue) and 2G (purple) stars. Vertical dashed lines mark the core radius (0.36 arcmin) and half-light radius (3.17 arcmin). Bottom: Radial variation of the rotational support to velocity dispersion ratio for both populations. Horizontal bars indicate bin widths.
    }
    \label{fig:pmt}
\end{figure}

In the top panel of Figure \ref{fig:pmt}, we compared the mean radial profiles of $\mu_\mathrm{T}$ for 1G (blue points) and 2G (purple squares) stars, given that $\mu_\mathrm{T}$ serves as a tracer of rotation. For this analysis, we considered Gaia proper motions matched with the photometric catalog of \cite{lee2022a}, which covers a wide radial range, since \textit{HST} and \textit{JWST} proper motions are relative and do not display signatures of rotation. We binned the data considering equal-frequency bins, with a total of 1712 2G stars and 868 1G stars. Any discrepancies between these profiles could potentially indicate distinct rotational characteristics between the two populations. We did not observe significant differences between the profiles of 1G and 2G stars (see Section \ref{sec:kinpro} for a discussion on the velocity dispersion profiles). The bottom panel shows the radial profile of $\mu_\mathrm{T}(\sigma_\mathrm{R}^2+\sigma_\mathrm{T}^2)^{-1/2}$ for 1G and 2G stars, quantifying the ratio of rotational support to random motion pressure. From this analysis, differences in the rotational behavior of 1G and 2G stars become more pronounced, especially in the cluster's outer regions. Recent studies \citep{leitinger2025, dalessandro2024} have further explored the rotational kinematics of multiple populations, revealing that 2G stars tend to rotate more rapidly than 1G in 47 Tucanae \citep[see also][for a general study on the rotation of 47 Tucanae]{bellini2017a}.

\subsection{Skewness}
\label{sec:skew}

We investigated the skewness of the $\mu_T$ versus $\mu_R$ distribution for 1G and 2G stars. 
Skewness quantifies the asymmetry of a probability distribution about its mean. We measured skewness using the Fisher-Pearson coefficient (implemented via the skew function in SciPy), which is the third standardized moment of the distribution.
A positive skewness indicates a
distribution with an elongated tail toward values larger than the mean, while negative skewness indicates a tail extending toward values smaller than the mean.
The top right panel of Figure \ref{fig:pms} clearly illustrates that the distribution exhibits skewness in the $\mu_T$ values. Indeed, a skew in 47 Tucanae was previously reported by \cite{heyl2017}, who analyzed skewness across all stars. In this study, we focused on the skewness of 1G and 2G stars separately to determine whether their orbital properties differ.

\begin{figure}
    \centering
    \includegraphics[width=0.95\linewidth]{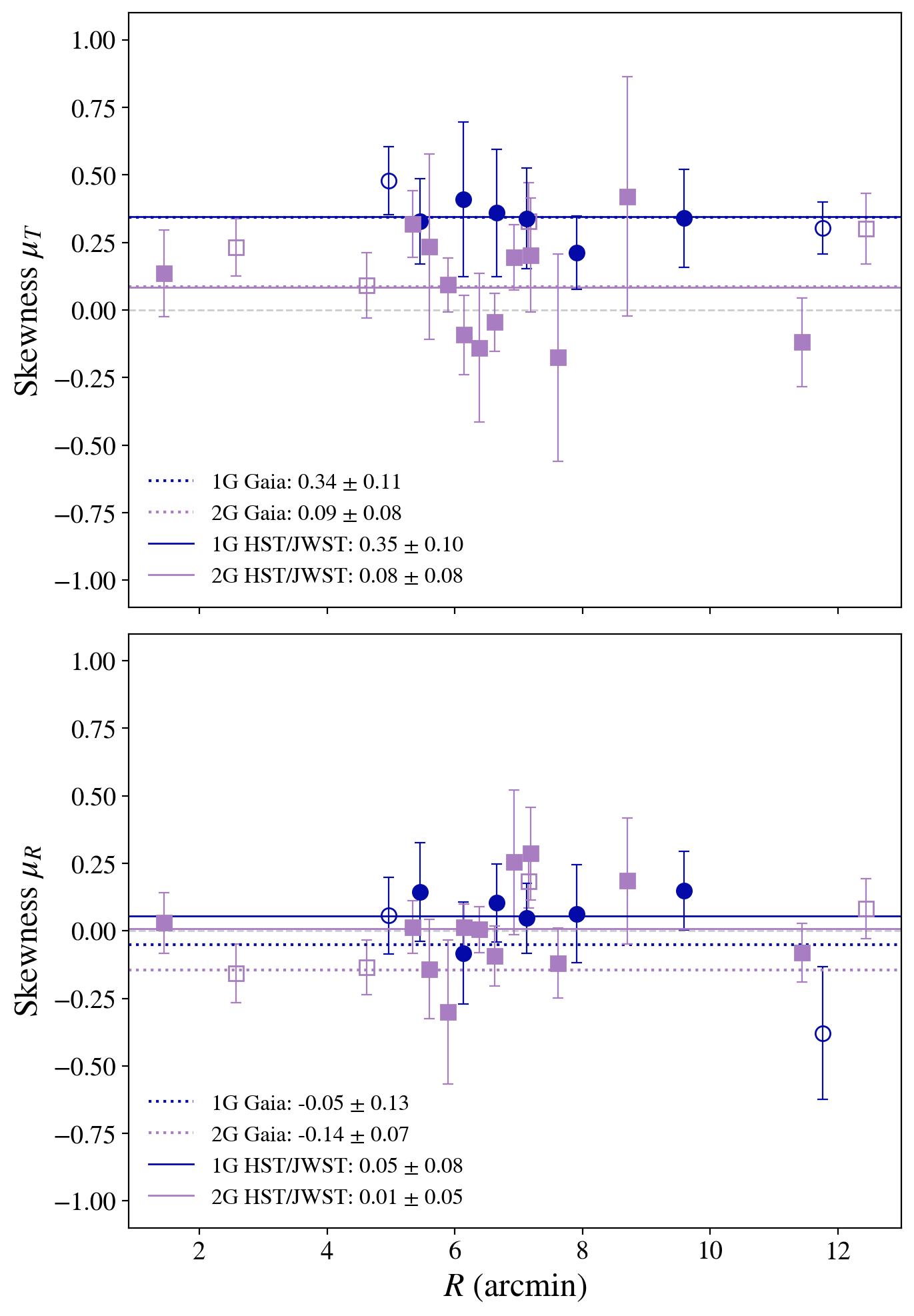}
    \caption{Radial variation of skewness in tangential (top) and radial (bottom) proper motions for 1G (blue) and 2G (purple) stars. Filled symbols represent \textit{HST}/\textit{JWST} data, while open symbols show Gaia measurements. Dashed (Gaia) and solid (\textit{HST}/\textit{JWST}) lines indicate the mean skewness for each population.}
    \label{fig:skew}
\end{figure}

Figure \ref{fig:skew} presents the skewness in $\mu_T$ for 1G stars (depicted as blue points) and 2G stars (shown as purple points) in the top panel. 
To compute the skewness, we divided the data into equal star count bins based on radial distance. For each bin, we used bootstrap resampling to estimate the skewness. Specifically, we generated random samples from the data by resampling with replacement and assuming normal distributions based on the observed data and their associated errors. We then computed the skewness for each bootstrap sample. The mean skewness value for each bin is shown, with the standard deviation indicating the uncertainty.
Open symbols refer to Gaia data, while filled symbols represent \textit{HST}/\textit{JWST} data.
Dashed (Gaia) and solid (\textit{HST}/\textit{JWST}) lines correspond to the mean skewness for each population determined by calculating the skewness using the proper motions of all stars of each population at all radii.
We display 1G measurements in blue, and 2G in purple.
The bottom panel displays analogous results for $\mu_R$. On average, both 1G and 2G stars exhibit approximately zero skewness in $\mu_R$. However, in the tangential component, we observed an average skewness significantly higher for 1G stars compared to 2G stars. These results are consistent with the measurements of \cite{heyl2017} for skewness of all stars combined in 47\,Tucanae. 

In studies of galaxy kinematics based on line-of-sight velocities, skewed velocity distribution profiles are often linked to internal rotation \citep[see, e.g.,][]{vandermarel1993,vandesande2017}. However, in our investigation — based on proper motion measurements — the interpretation of differences (or lack thereof) in skewness between the tangential and radial velocity distributions is less straightforward. Notably, the differences in tangential velocity skewness between the two stellar generations do not correspond to significant differences in their rotational velocities.

Other dynamical effects could also play a role in shaping the velocity distribution profile. In particular, the complex interaction between internal dynamical processes and the Galactic tidal field along with the escape of stars could affect the skewness of the cluster's velocity distribution. Any effect related to the external tidal field, however, would presumably affect more significantly the cluster's outermost regions and lead to a radial variation in skewness which contrasts with the observed approximately constant skewness for 1G stars. Therefore, our results do not provide strong evidence that tidal stripping plays a significant role in explaining the observed skewness of 1G and 2G stars. Further exploration through dedicated N-body simulations is needed to better understand how the interplay between the initial dynamical properties of the two populations and the effects of various dynamical processes influences the evolution of the velocity distribution skewness.
Nevertheless, the velocity distribution skewness measured in our analysis provides additional evidence of kinematic differences between the two stellar generations and sets a new constraint for models of the formation and dynamical evolution of multiple populations.

\subsection{Velocity dispersion}

Prior to analyzing the velocity dispersion, anisotropy profiles, and energy equipartition, we removed the rotational component from the Gaia proper motions for consistency with the relative proper motions from HST and JWST. This was achieved by fitting sinusoidal functions to the proper motions in the $x$ and $y$ directions as a function of stellar position angle. The residuals from these fits were then adopted as the de-rotated proper motions in the respective directions.

\begin{figure}
    \centering
    \includegraphics[width=7cm]{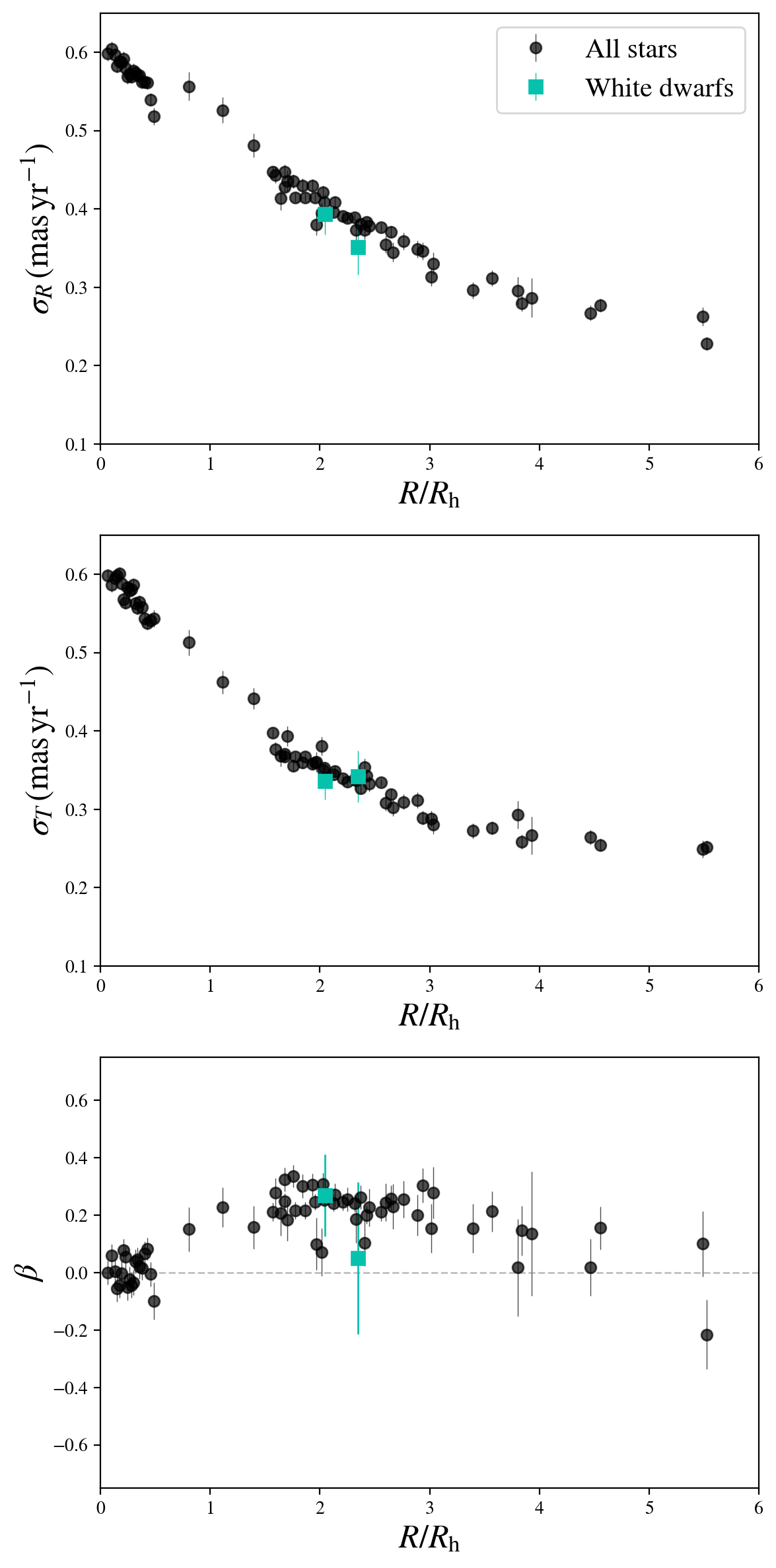}
    \caption{Kinematic profiles showing radial velocity dispersion (top), tangential velocity dispersion (middle), and anisotropy $\beta = 1 - \sigma_T^2/\sigma_R^2$ (bottom) for all stars. White dwarfs from field A are highlighted in turquoise.}
    \label{fig:wd}
\end{figure}

To determine the velocity dispersion, we divided the field of view into a series of concentric annular regions,
each including approximately the same number of stars. For each annulus we determined the dispersion from either Gaia DR3, \textit{JWST} or \textit{HST} radial and tangential velocity components by maximizing the log-likelihood function described in Equation 1 of \cite{libralato2022a}, for \textit{HST}/\textit{JWST}, and \citet{bianchini2018b}\footnote{updated including the covariance term as in \citet{sollima2019}} for Gaia. We determined the uncertainties in the velocity dispersions using the Markov Chain Monte Carlo algorithm emcee \citep{foreman2013}. Figure \ref{fig:wd} illustrates the velocity dispersion and anisotropy (as defined in Equation \ref{eq:ani}) profiles for all stars in our dataset. Our observations include proper motions of white dwarf stars in Field A, that are highlighted in turquoise.

\subsection{Velocity anisotropy model} \label{subsection:animodel}

Following the results of simulations of the evolution of multiple-population GCs, we model the 2D radial variation of the velocity anisotropy using the following analytical expression \citep[see][]{dalessandro2024,aros2025}.

\begin{equation}
\beta(R) = 1-\frac{\sigma_\mathrm{T}^2}{\sigma_\mathrm{R}^2} = \frac{\beta_{\infty}R^2}{R_a^2+R^2}\left(1-\frac{R}{R_\mathrm{t}}\right)
\label{eq:ani}
\end{equation}

where $\sigma_\mathrm{T}$ and $\sigma_\mathrm{R}$ are the velocity dispersions in the tangential and radial directions, respectively, $R_\mathrm{a}$ is the anisotropy radius, $\beta_{\infty}$ is the anisotropy at $R \gg R_\mathrm{a}$, and $R_\mathrm{t}$ is the truncation radius, defined as the radius for which the velocity becomes isotropic again \citep[the model assumes isotropy at the cluster center, and increasing anisotropy for $R > R_\mathrm{a}$; see][for more details]{dalessandro2024,aros2025}. This parametrization is a modification of the Osipkov-Merrit model \citep{osipkov1979,merritt1985},  allowing a free parameter for the anisotropy (either radial or tangential velocity dispersion for $R > R_\mathrm{a}$).

To find the best-fit values for the model, we describe the radial velocity dispersion as in \cite{watkins2022}:

\begin{equation}
    \sigma_{R}(R) = c_\mathrm{0}+c_\mathrm{1}R+c_\mathrm{2}R^2+c_\mathrm{3}R^3
\end{equation}

where the coefficients should comply with the following conditions:

\begin{equation}
    c_\mathrm{0}+c_\mathrm{1}R+c_\mathrm{2}R^2+c_\mathrm{3}R^3 \geq 0
\end{equation}
\vspace{-1cm}
\begin{equation}
    c_\mathrm{1}+2c_\mathrm{2}R+3c_\mathrm{3}R^2 < 0.
\end{equation}

The anisotropy is related to the tangential velocity dispersion, expressed as

\begin{equation}
    \sigma_\mathrm{T}^2 = (1-\beta(R | R_\mathrm{a},\beta_{\infty},R_\mathrm{t}))\sigma_\mathrm{R}^2.
\end{equation}

With the observed radial and tangential proper motions, we performed a Monte Carlo Markov-Chain fit to get the best-fit values for $c_\mathrm{0}$, $c_\mathrm{1}$, $c_\mathrm{2}$, $c_\mathrm{3}$, $R_\mathrm{a}$, $\beta_{\infty}$ and $R_\mathrm{t}$. 

\begin{figure*}
    \centering
    \includegraphics[width=16.25cm]{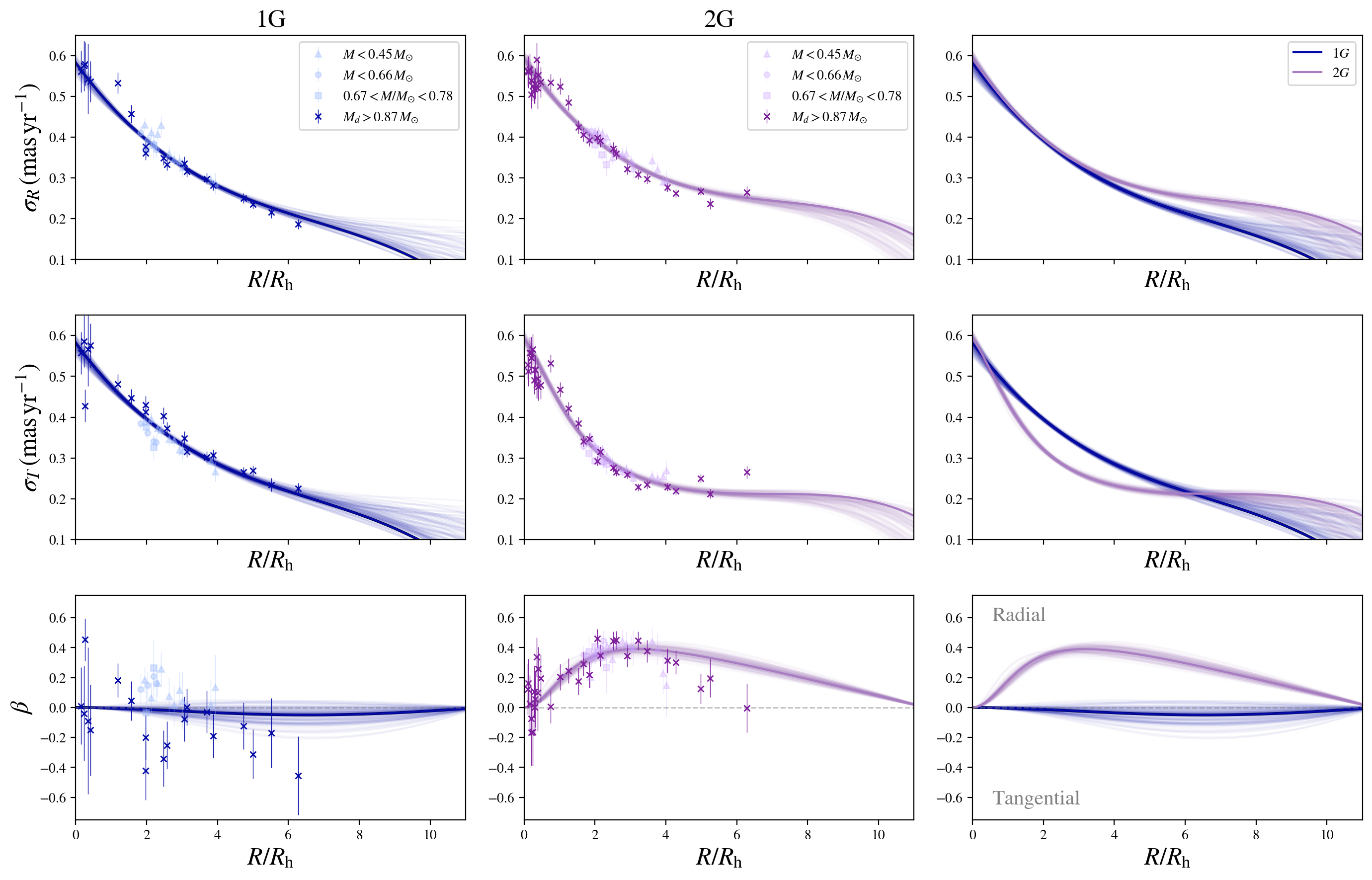}
    \caption{Radial profiles of the velocity dispersion in the radial (top panels) and tangential (middle panels) directions. The bottom panels show the anisotropy profiles, with $\beta = 1 - (\sigma_T/\sigma_R)^2$. 1G observations are represented with blue colors, and 2G with purple colors. The dark blue and purple lines indicate the best-fit models, as discussed in Section \ref{subsection:animodel}. The light blue and purple lines represent additional model realizations based on sampled parameter sets. The radial coordinate is normalized to the half-light radius, $R_h = 3.17$ arcmin.}
    \label{fig:ani}
\end{figure*}

\begin{figure*}
    \centering
    \includegraphics[width=16.2cm]{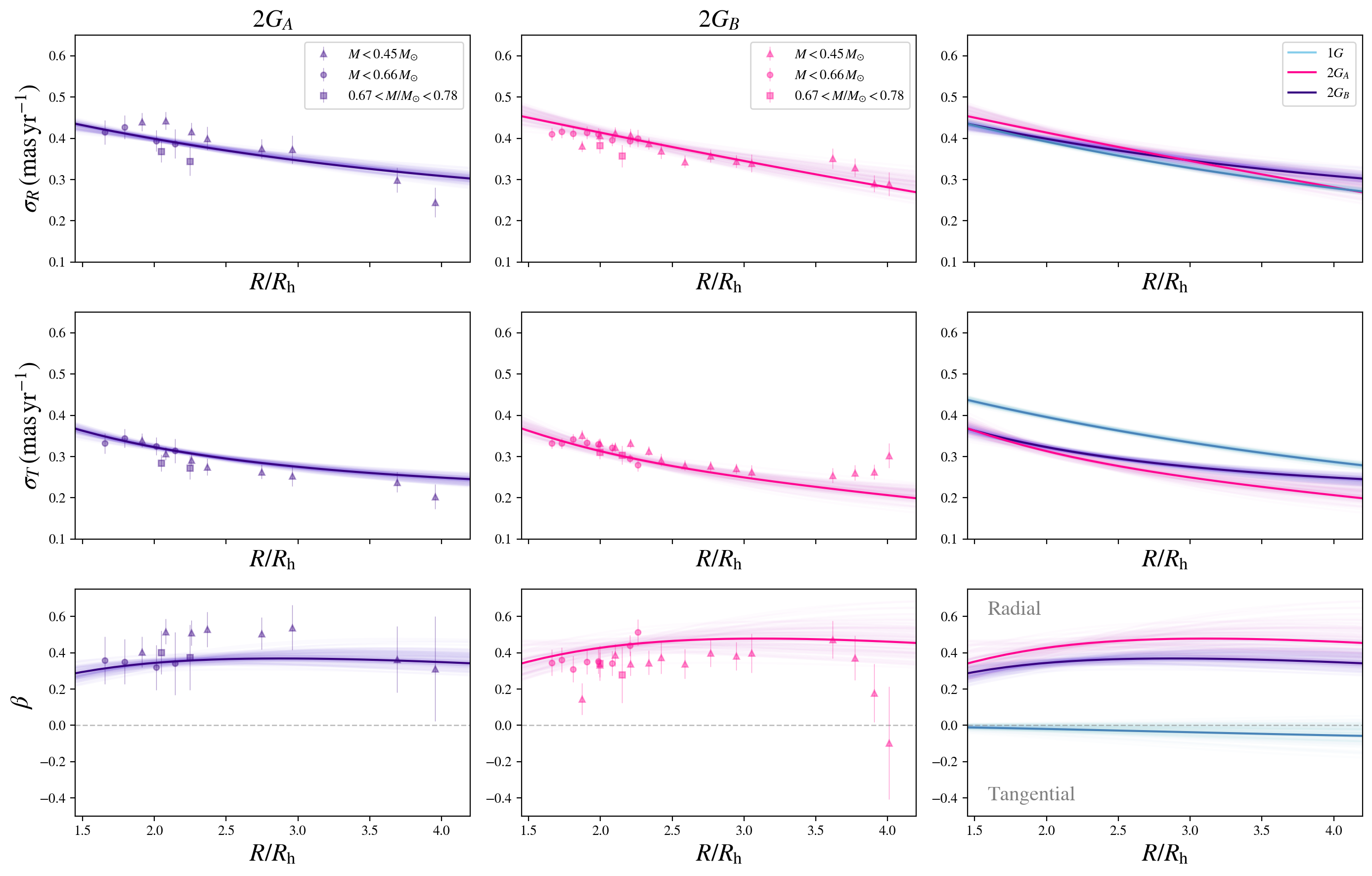}
    \caption{Similar to Figure \ref{fig:ani}, but for $2G_A$ and $2G_B$ subpopulation stars (pink and purple points, respectively, with corresponding colored lines), over a more limited radial range. In the third column, the fit for $1G$ stars is displayed for comparison.}
    \label{fig:2gab}
\end{figure*}

\subsection{Anisotropy of the multiple populations} \label{sec:kinpro}

Figure \ref{fig:ani} shows the velocity dispersion profiles in the radial (top panels) and tangential (middle panels) directions, along with the anisotropy profiles (bottom panels). The best-fit model profiles
are indicated by the dark solid lines. Triangles correspond to data from fields A (lower-MS), B and C, circles to field D, squares to field A (upper-MS), and crosses to the inner field, JWL, Stetson, and Gaia XP data. The first column shows 1G stars (in blue), while 2G stars are shown in purple in the second column. For better comparison, the third column displays the comparison between the 1G and 2G best fit profiles, obtained as described in section \ref{subsection:animodel}. The radial distances are expressed in units of half-light radius ($R_\mathrm{h}$), which is 3.17 arcminutes for 47\,Tucanae \citep[][2010 version]{harris1996a}. 

1G and 2G stars present similar velocity dispersions in the radial direction. In the tangential direction, 1G stars present higher velocity dispersion than 2G stars, except in the very center and outer regions (approximately $\geq 5$ $R_h$), where they are very similar. This difference translates into different anisotropy profiles. While 1G stars' motions are isotropic, 2G stars present radial anisotropy. Overall, Gaia and \textit{HST}/\textit{JWST} profiles are in good agreement within the uncertainties. However, since Gaia targets giant stars, while \textit{HST} and \textit{JWST} proper motions are for MS stars, small differences are expected due to energy equipartition \citep[see e.g.][]{pavlik2024}.

We further examine 2G stars, which can be divided into two subpopulations, $2G_A$ and $2G_B$, with the latter representing the chemically extreme group of 2G stars located in the upper-left regions of the ChMs. The selection of ChMs for which we could identify these subpopulations can be visualized among the ChMs of Figure \ref{fig:chms}. Similarly to our analysis of 1G and 2G kinematic profiles, we compared the profiles of these subpopulations, as shown in Figure \ref{fig:2gab}. The velocity dispersion profiles of $2G_A$ and $2G_B$ stars exhibit remarkable similarity, with only minor differences that manifest in their anisotropy profiles. Notably, the more chemically extreme $2G_B$ population shows a marginally higher radial anisotropy profile compared to $2G_A$ stars.

\subsection{Energy equipartition} \label{sec:equi}

In this section, we describe the evolution toward energy equipartition of the multiple populations of 47\,Tucanae. 
Since the degree of energy equipartition is expected to change with radius,
typically increasing toward the center \citep{trenti2013,pavlik2021,pavlik2022,watkins2022,aros2023,pavlik2024}, we analyze the mass dependence of the velocity dispersion in different radial regions corresponding to the fields described in Figure \ref{fig:footprint}.

\begin{figure*}
    \centering
    \includegraphics[width=0.245\textwidth]{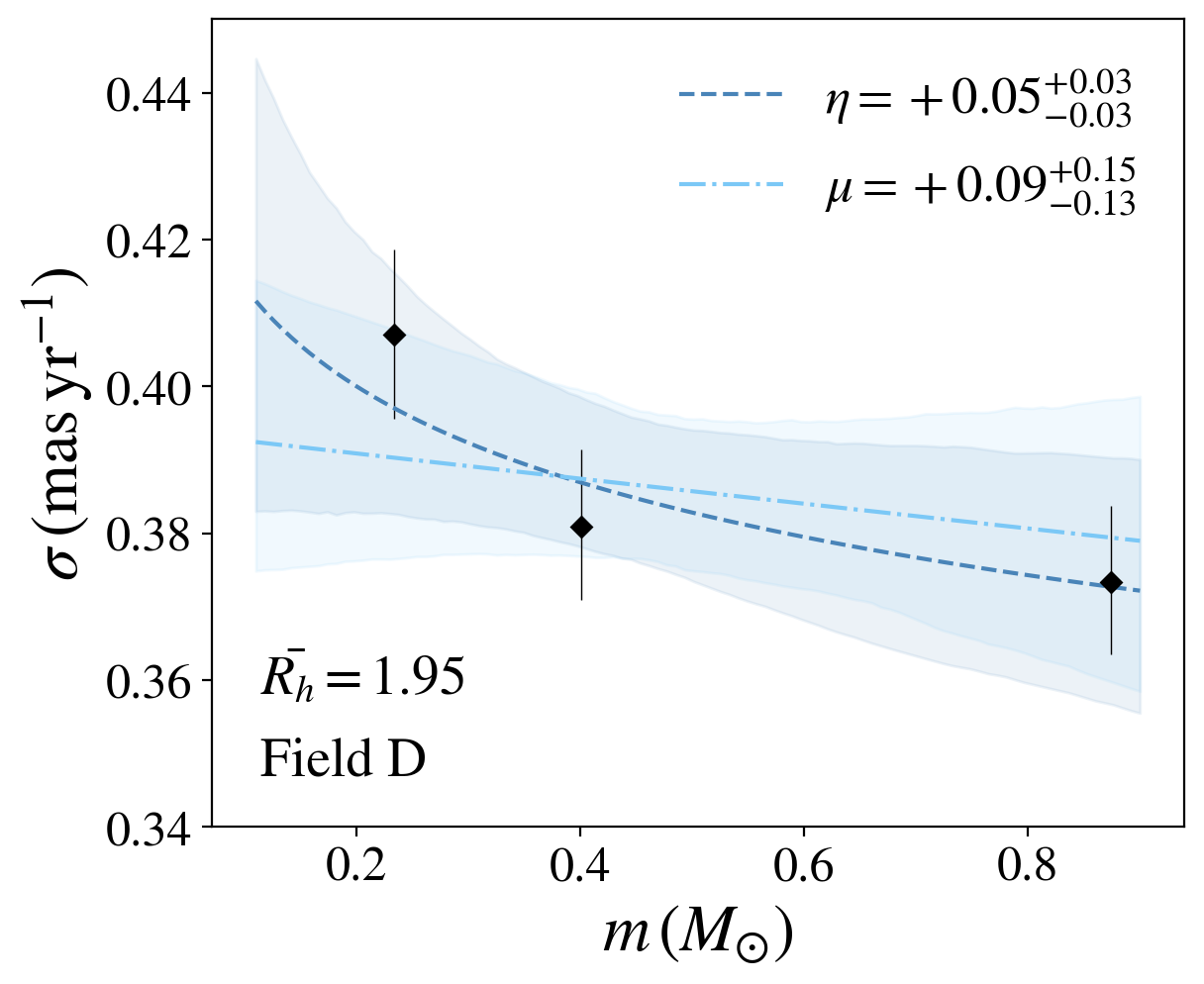}
    \hfill
    \includegraphics[width=0.245\textwidth]{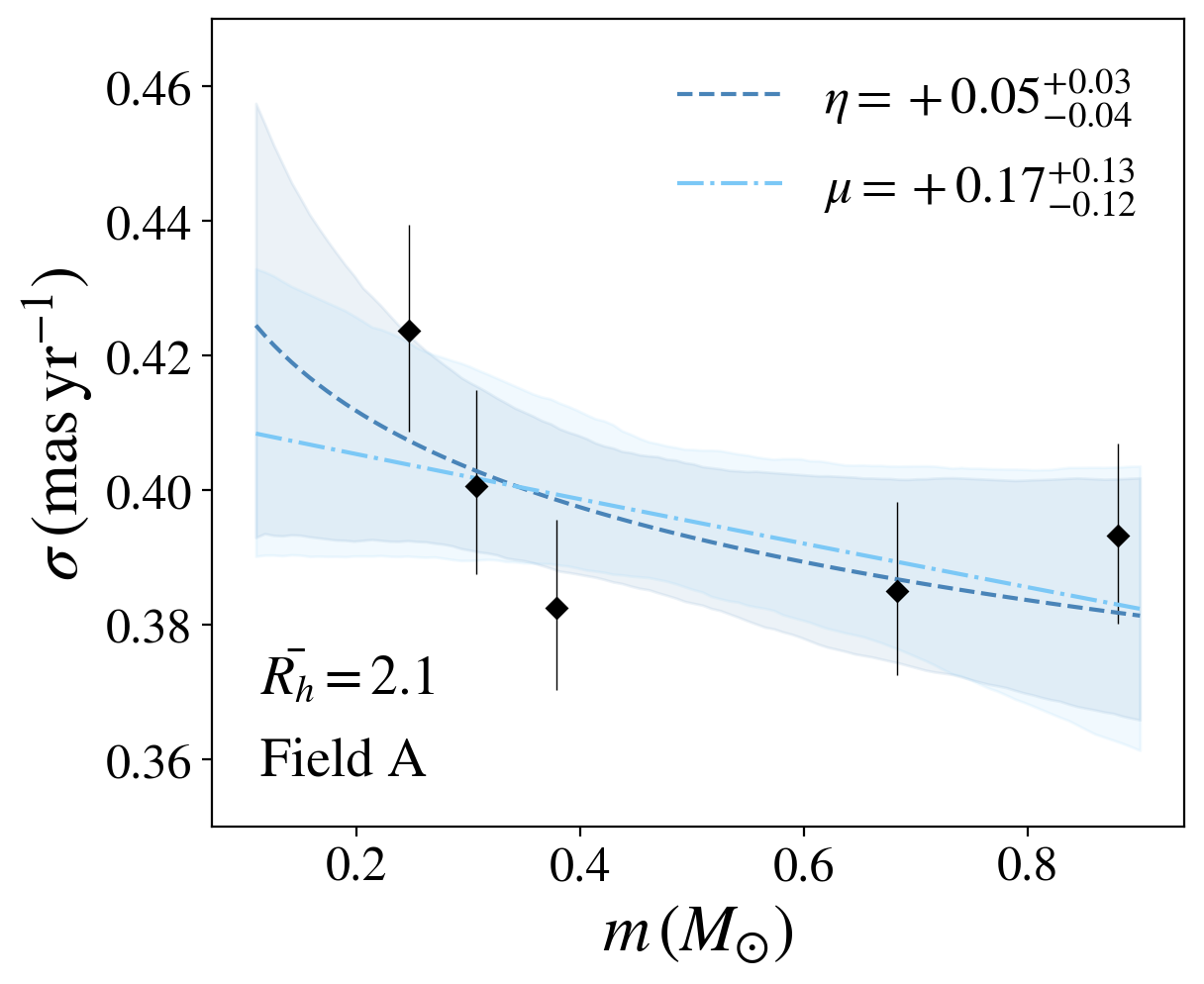}
    \hfill
    \includegraphics[width=0.245\textwidth]{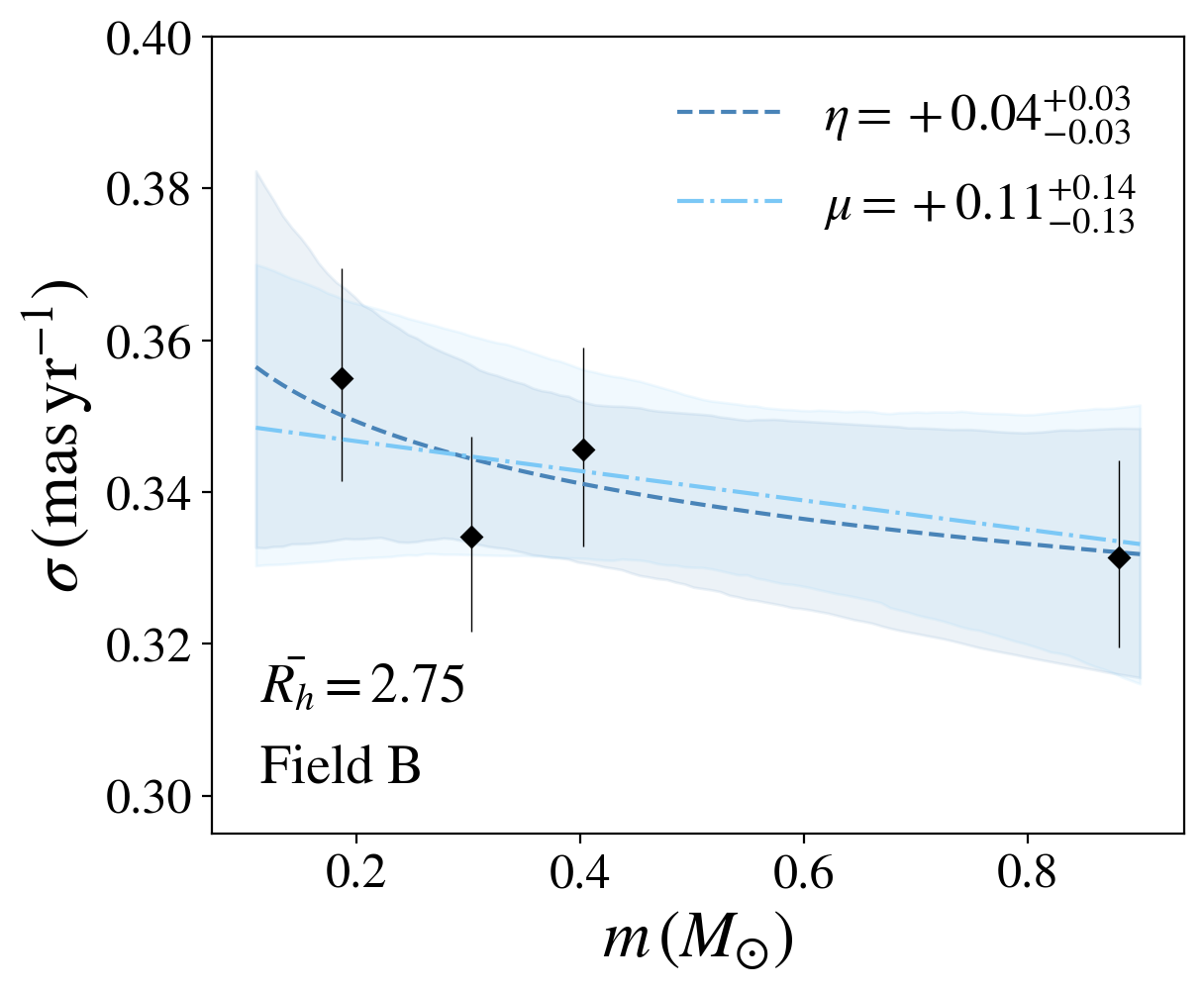}
    \hfill
    \includegraphics[width=0.245\textwidth]{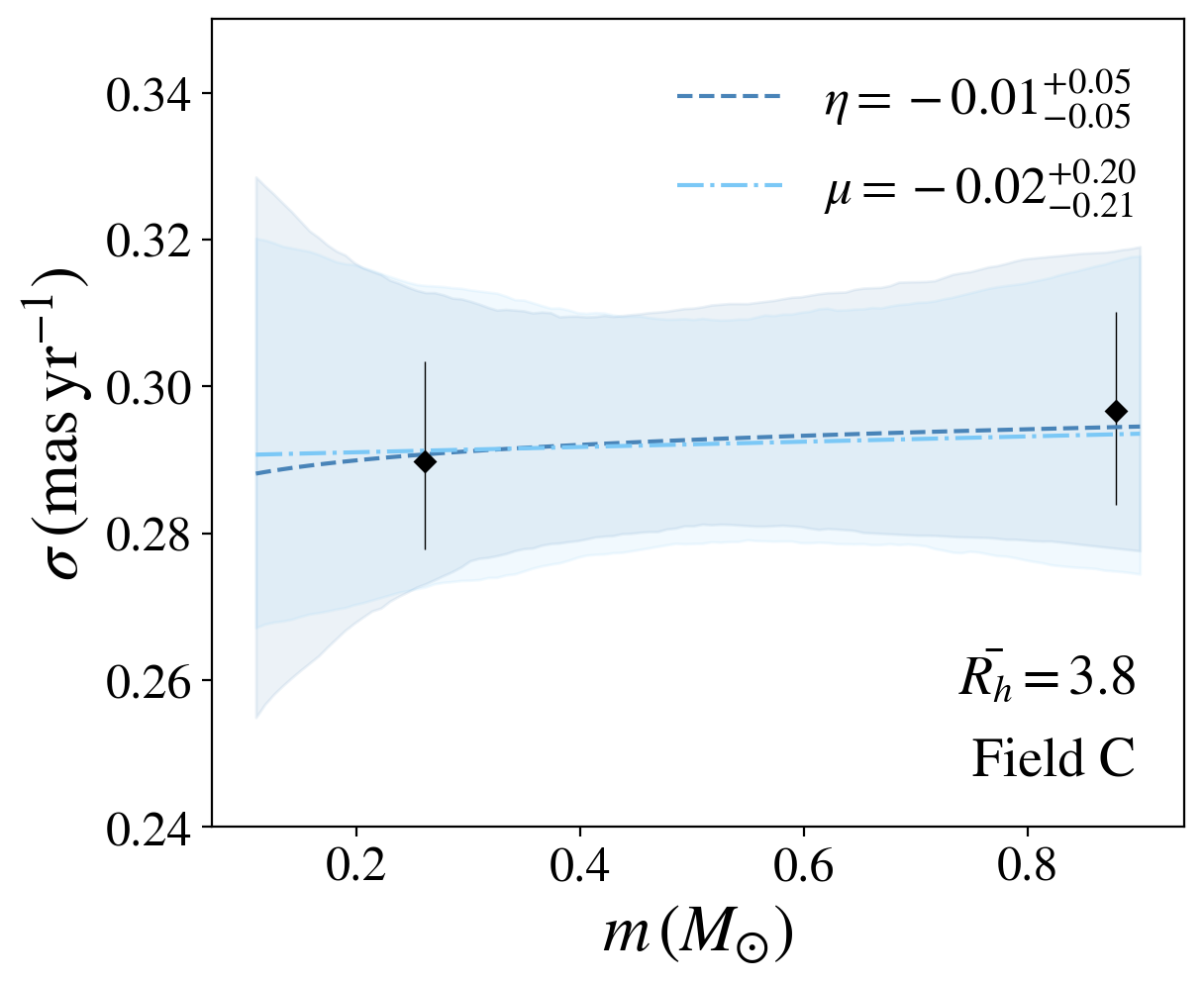}
    
    \includegraphics[width=0.245\textwidth]{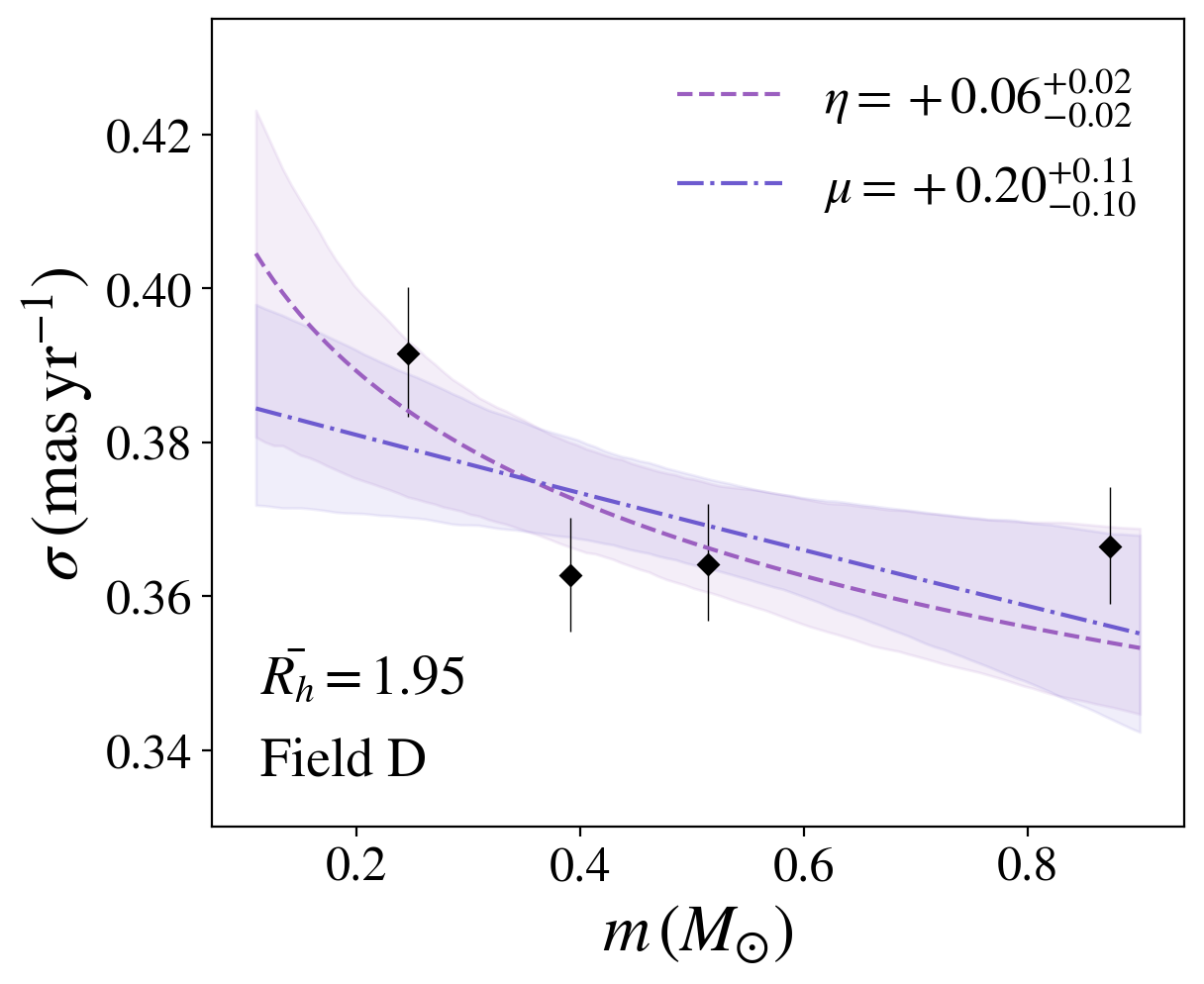}
    \hfill
    \includegraphics[width=0.245\textwidth]{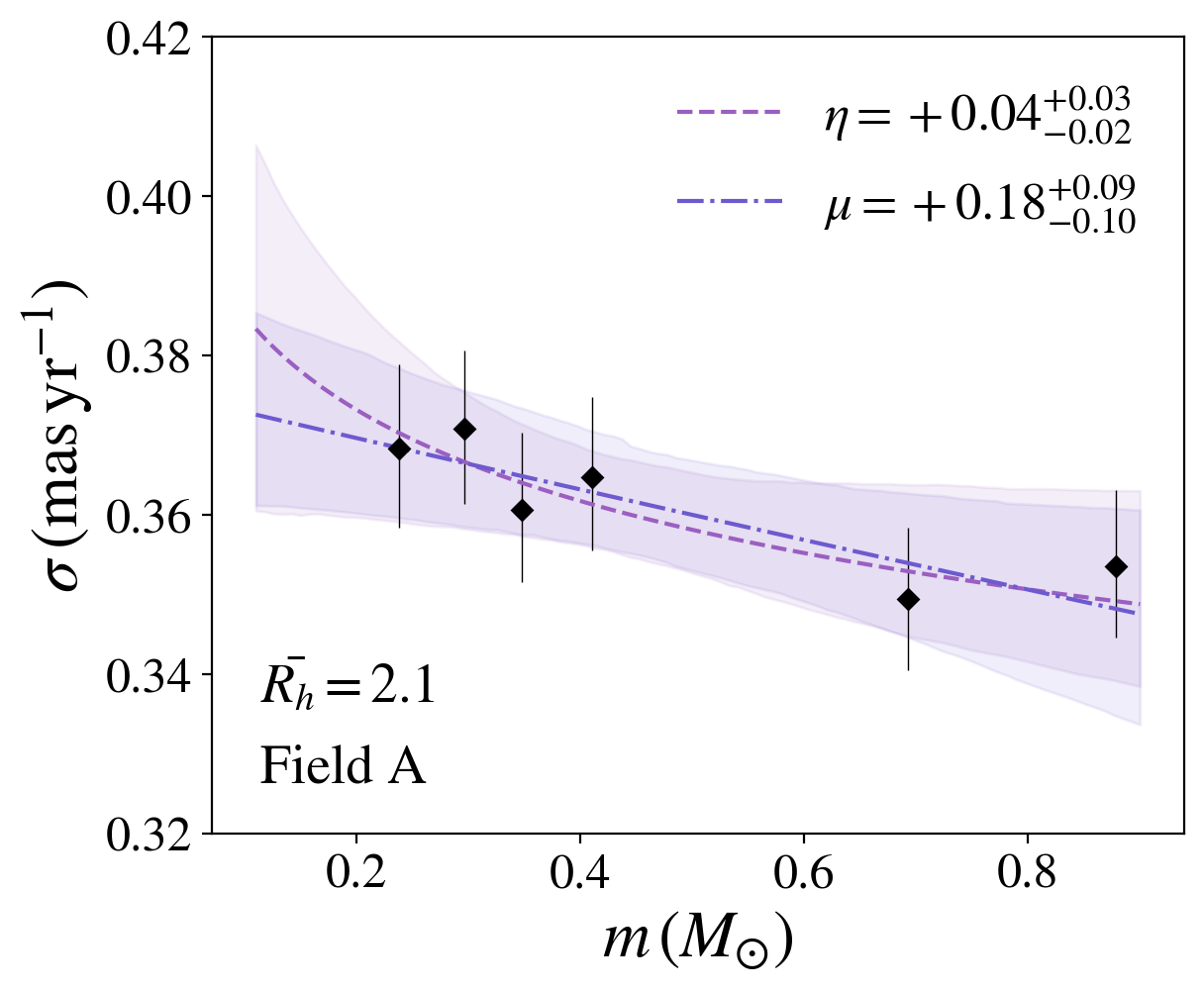}
    \hfill
    \includegraphics[width=0.245\textwidth]{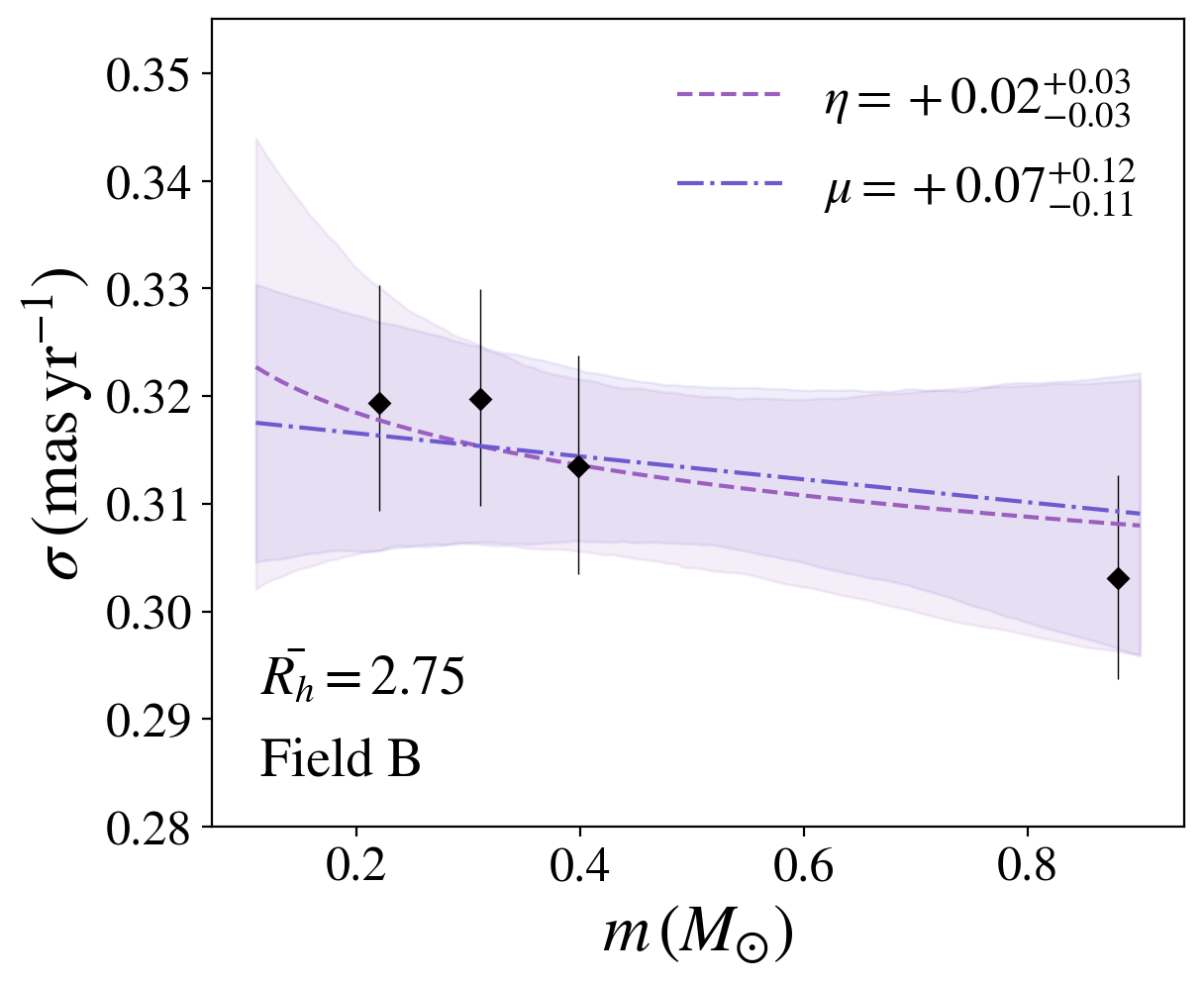}
    \hfill
    \includegraphics[width=0.245\textwidth]{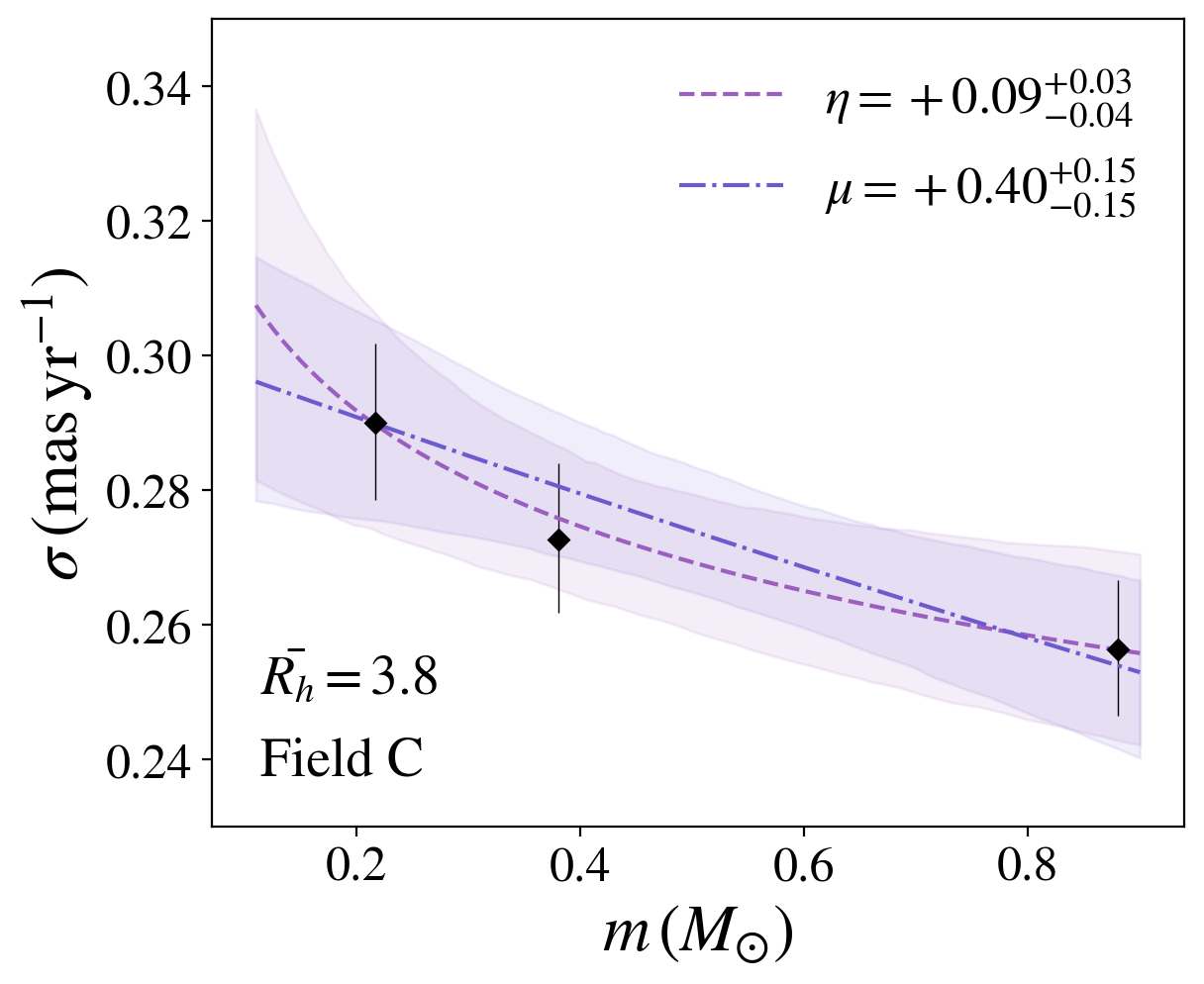}    
    \caption{Velocity dispersion as a function of stellar mass for 1G (top panels) and 2G (bottom panels) stars. The labels at the bottom of the plots indicate the average radius in units of half-light radius for stars in each field, and the name of the field. The best-fit values of $\eta$ and $\mu$ are indicated at the top right of each panel. Shaded regions represent the 5th to 95th percentile range of the models.}
    \label{fig:eqall}
\end{figure*}

\begin{figure*}[h]
    \centering
    \includegraphics[width=0.245\textwidth]{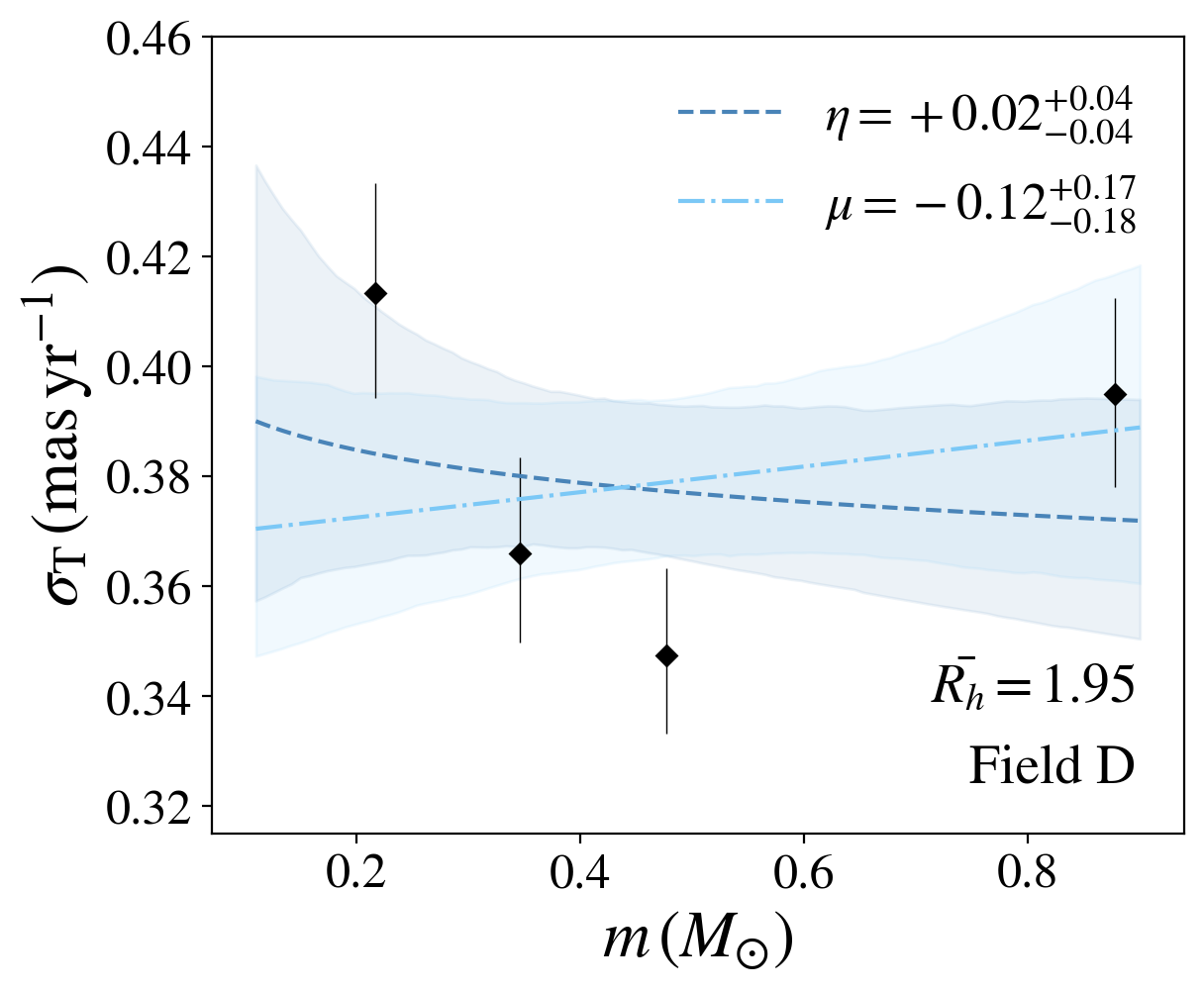}
    \hfill
    \includegraphics[width=0.2452\textwidth]{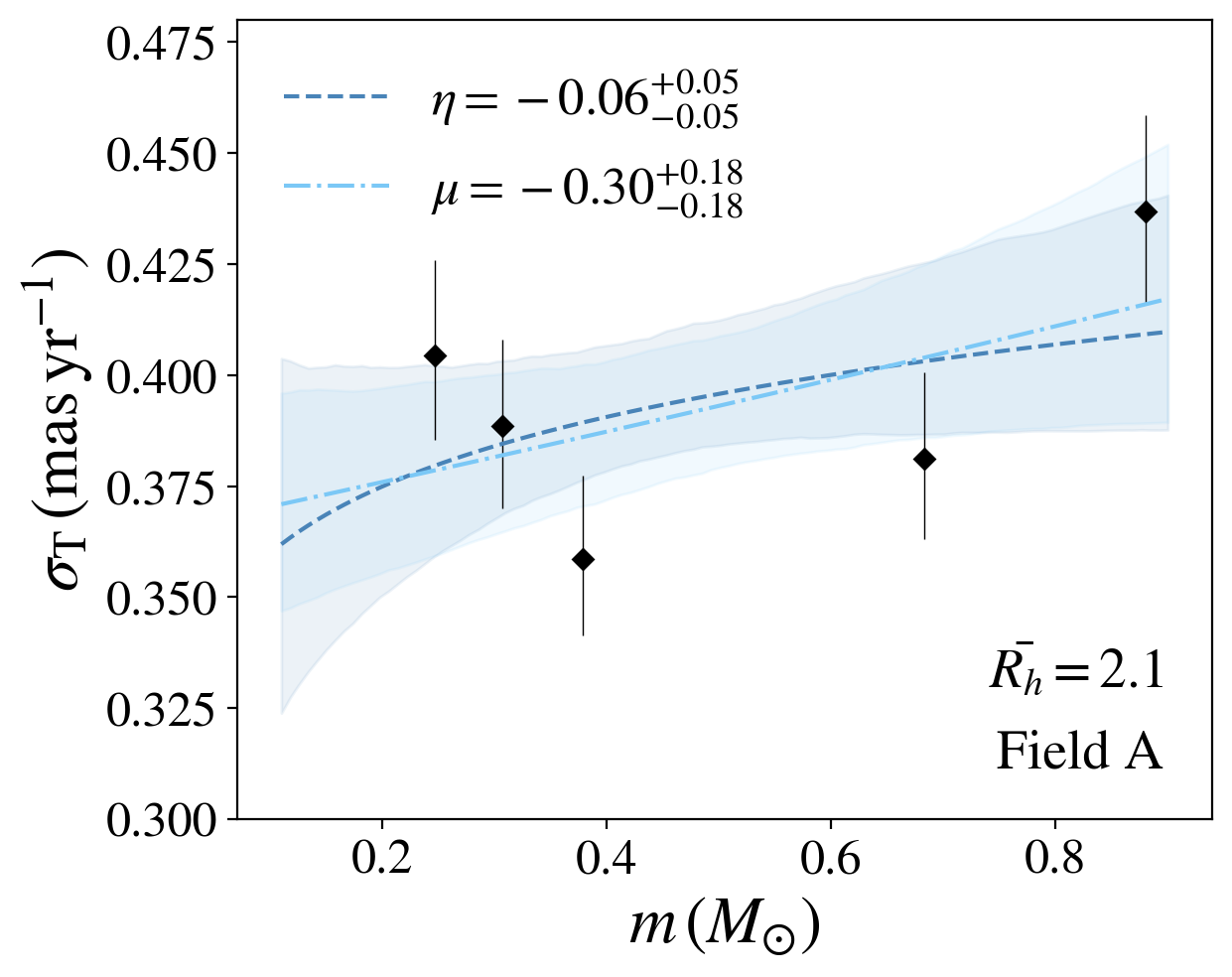}
    \hfill
    \includegraphics[width=0.245\textwidth]{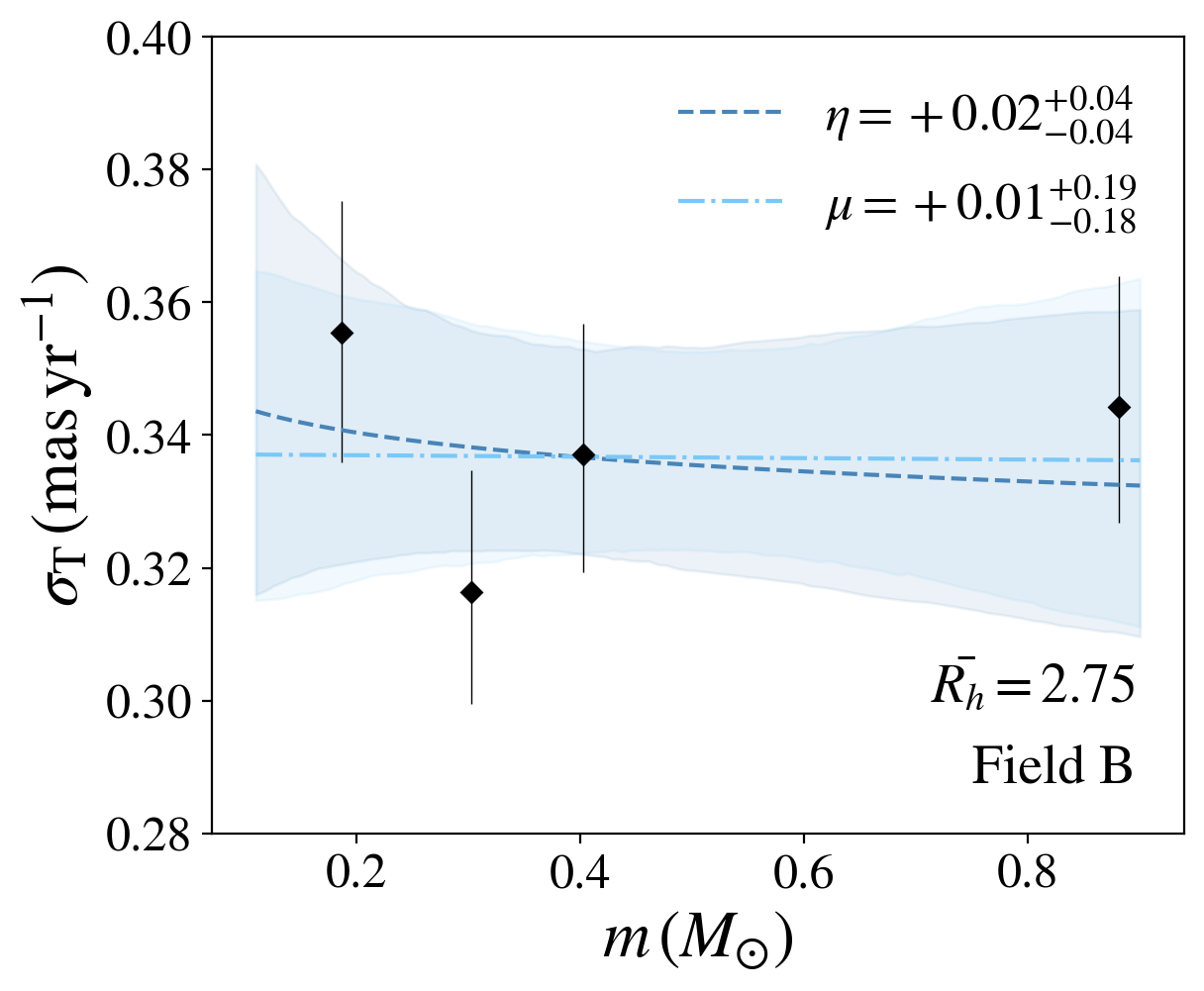}
    \hfill
    \includegraphics[width=0.245\textwidth]{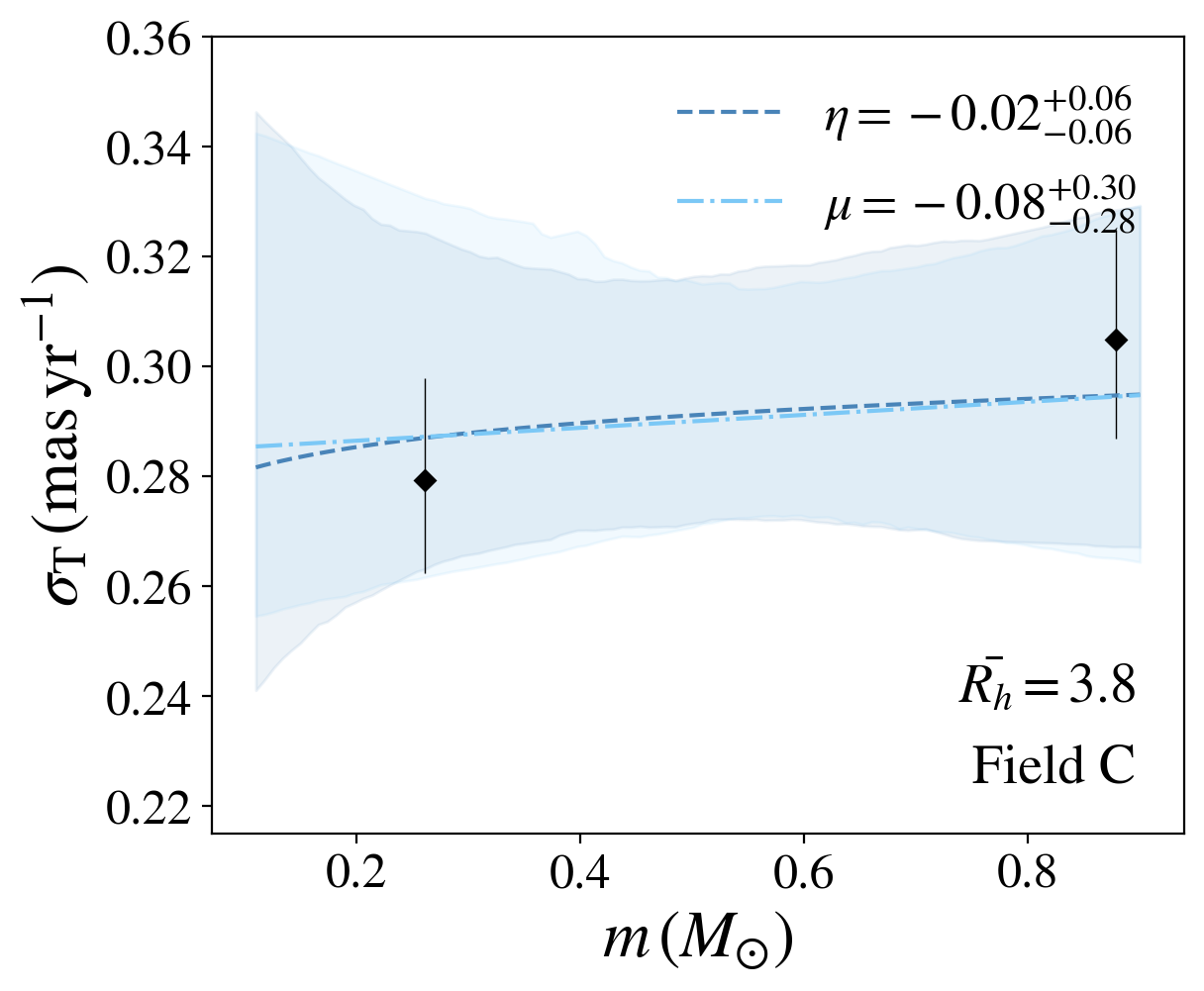}
    
    \includegraphics[width=0.245\textwidth]{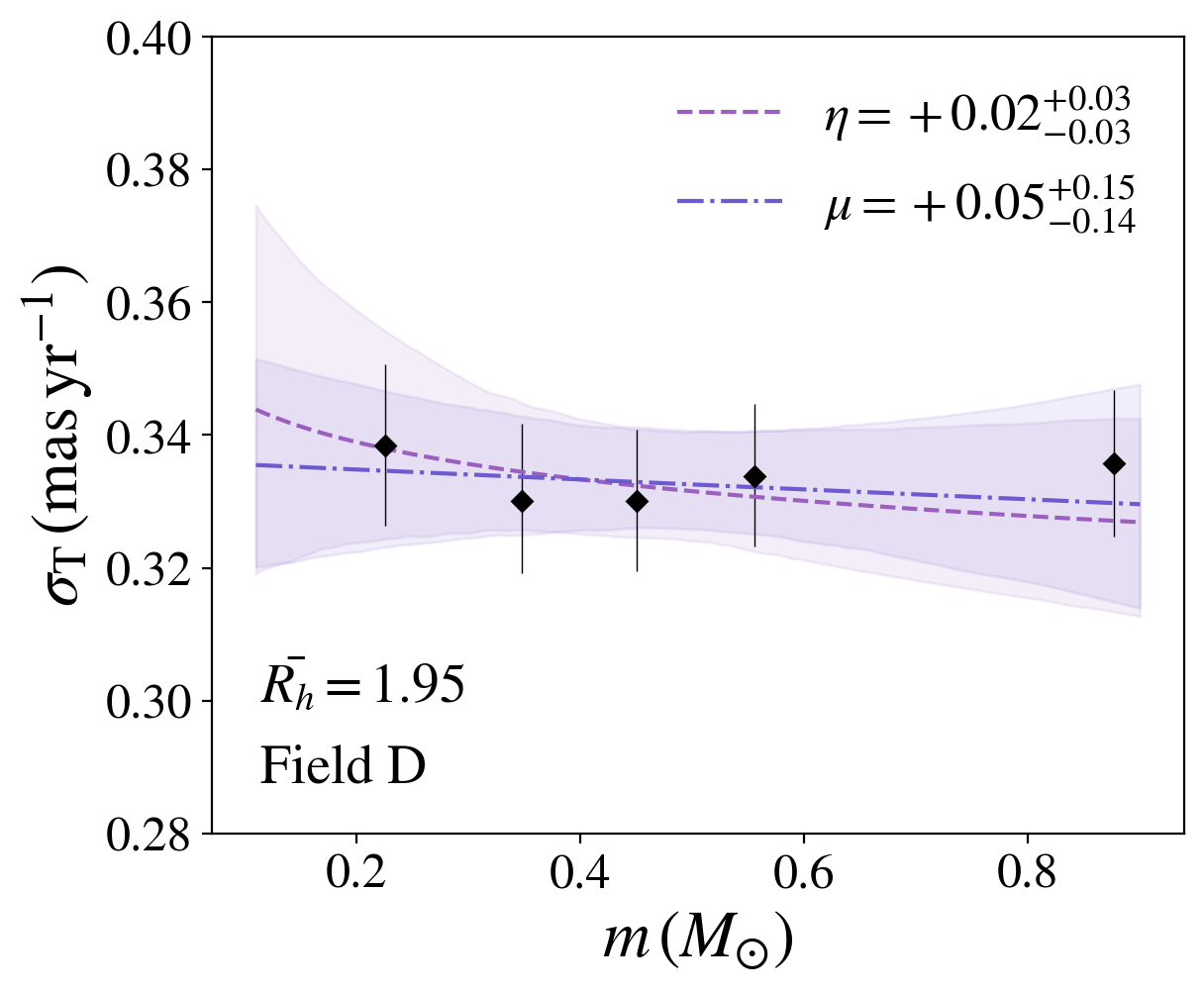}
    \hfill
    \includegraphics[width=0.245\textwidth]{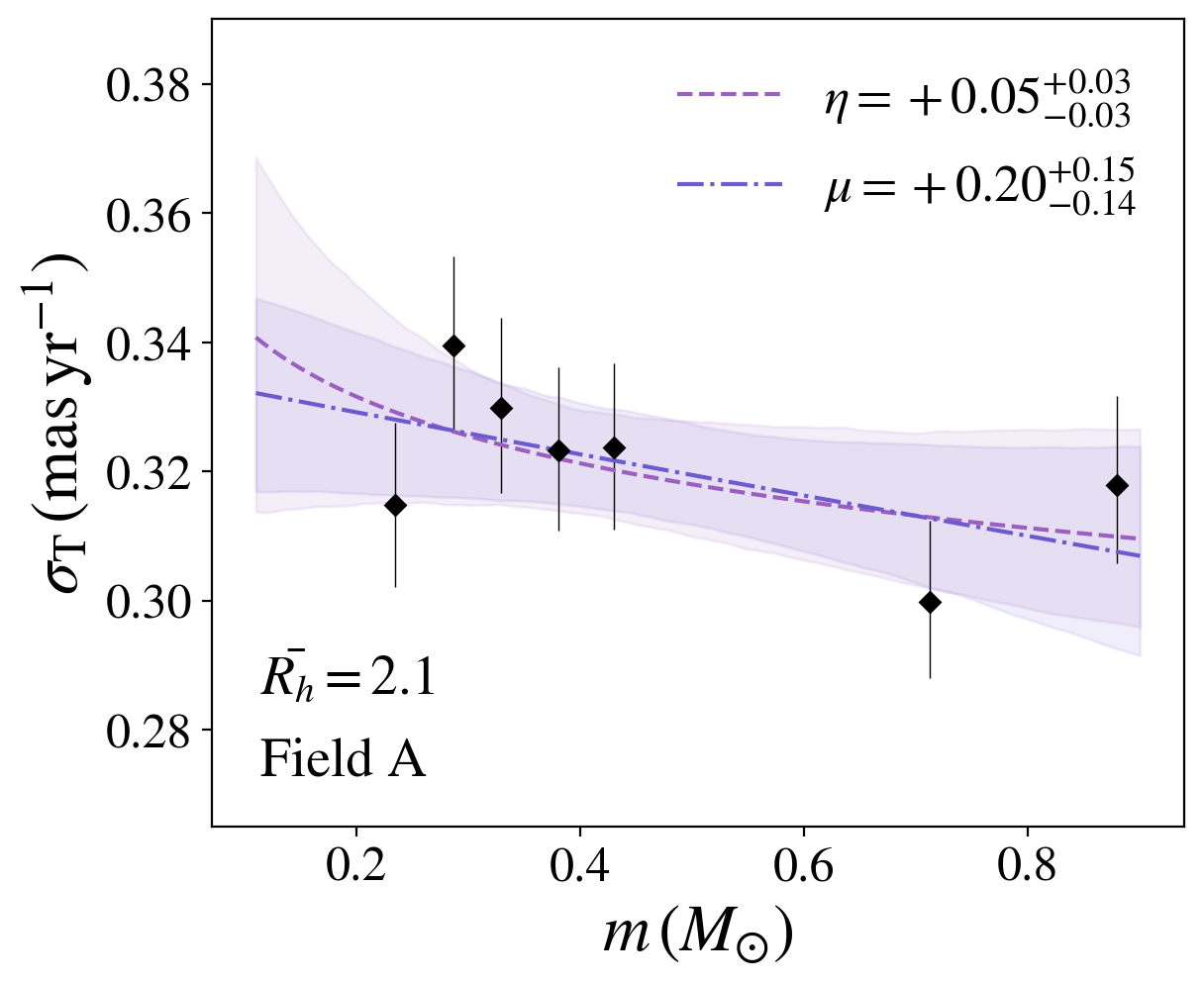}
    \hfill
    \includegraphics[width=0.245\textwidth]{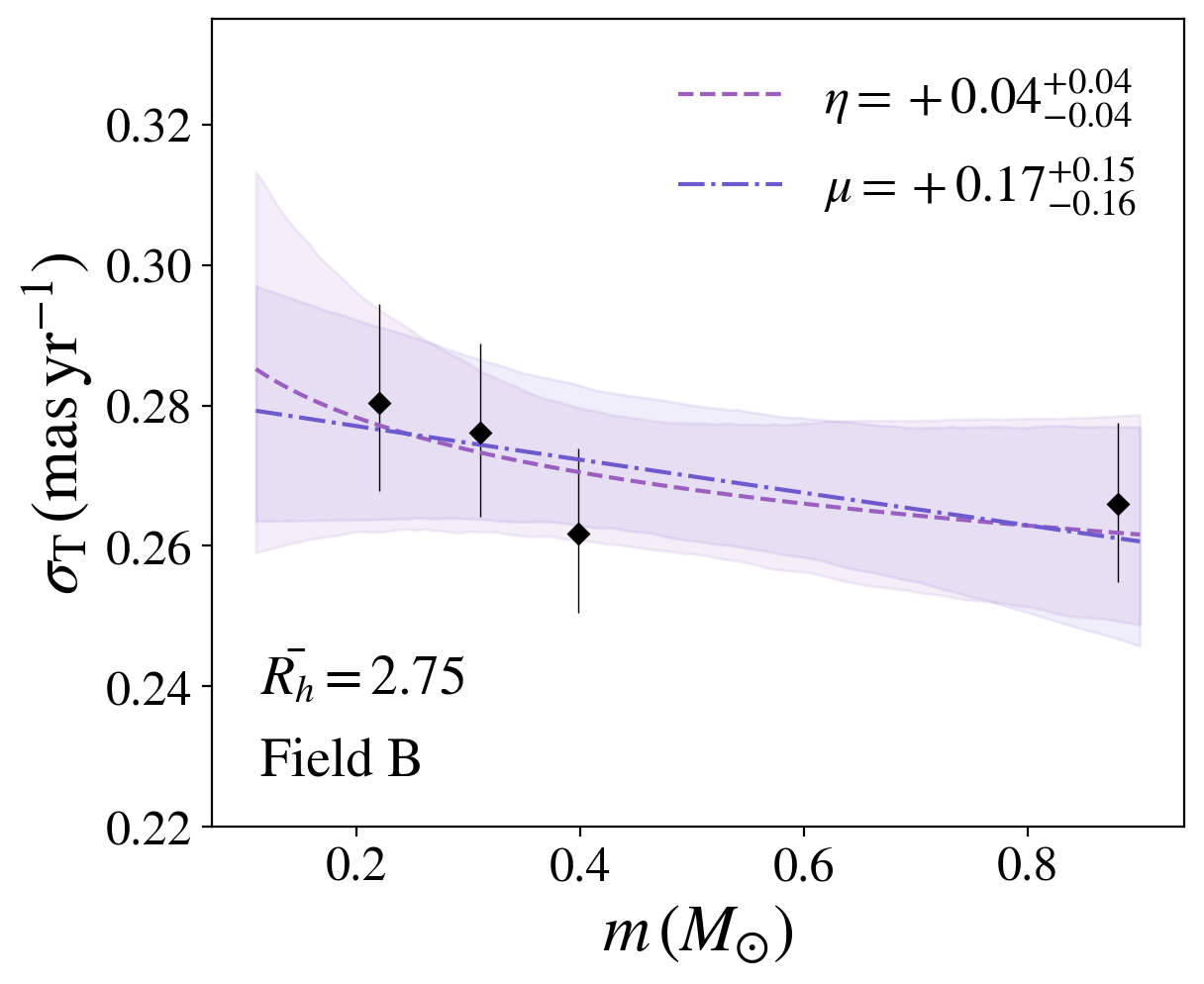}
    \hfill
    \includegraphics[width=0.245\textwidth]{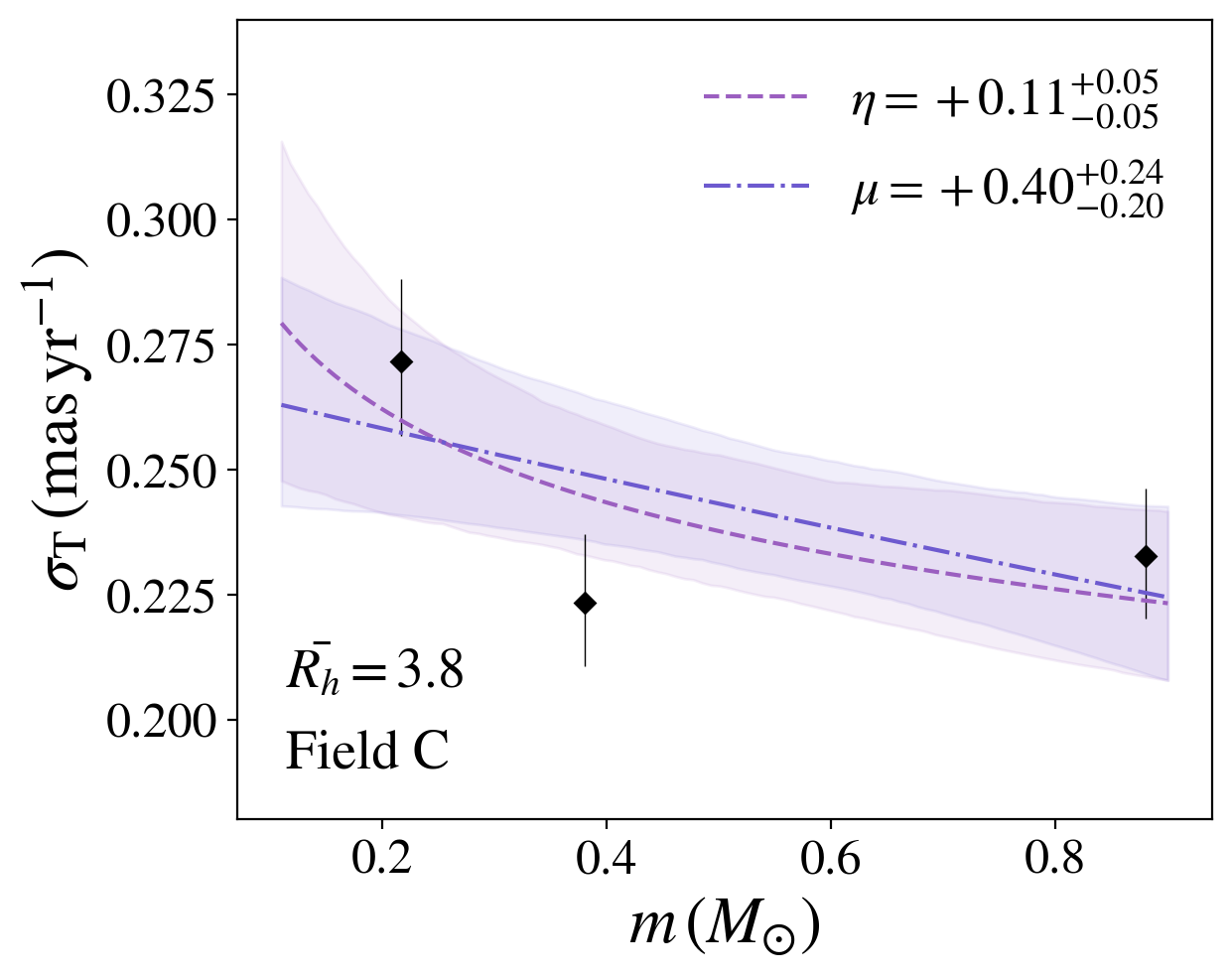}    
    \caption{Same as Figure \ref{fig:eqall}, but only considering the tangential component of the velocity dispersion.}
    \label{fig:eqtan}
\end{figure*}

We employ two models to describe the mass-dependent velocity dispersion. The first approach, following \cite{trenti2013}, is given by:

\begin{equation}
    \sigma(m) = \sigma_{0}\left(\frac{m}{m_{0}}\right)^{-\eta}
    \label{eq:classic}
\end{equation}

where $\sigma_{0}$ is the velocity dispersion for stars with mass $m = m_0$, and $m_0$ is a scale mass, defined here as $m_0 = 1 M_{\odot}$. The parameter $\eta$ quantifies the degree of energy equipartition, with $\eta = 0$ indicating no equipartition, $\eta = 0.5$ representing full equipartition, and negative values representing a trend opposite to an evolution toward energy equipartition — a configuration where higher-mass stars are faster than low-mass stars.

The second approach considers an equipartition mass, $m_{\mathrm{eq}}$, originally defined by \cite{bianchini2016} as follows:

\begin{equation}
\sigma(m) = 
    \begin{cases}
     \sigma_{0}\exp{(-0.5m/m_{\mathrm{eq}})}  & \mathrm{for \,} m \leq m_{\mathrm{eq}},\\
    \sigma_{\text{eq}}(m/m_{\mathrm{eq}})^{-0.5}  & \mathrm{for \,} m > m_{\mathrm{eq}}.
    \end{cases}
    \label{eq:bianchini}
\end{equation}

Stars with masses larger than $m_{\mathrm{eq}}$ are in full equipartition, while those with smaller masses are in partial equipartition. The value of $m_{\mathrm{}{eq}}$ serves as an indicator of the cluster's overall state of energy equipartition, with smaller values corresponding to clusters closer to full equipartition. $\sigma_\mathrm{eq}$ corresponds to the velocity dispersion at $m=m_{\rm eq}$ and is defined as $\sigma_{\mathrm{eq}} = \sigma_{0}\exp{(-0.5)}$. 

As in \cite{aros2023}, we define

\begin{equation}
    \mu = \frac{1}{m_{\mathrm{eq}}}
\end{equation}

to measure the degree of energy equipartition for Equation \ref{eq:bianchini}, as it allows for a continuous sampling of the parameter space in the fitting procedure, including negative values. $\mu = 0$ describes a system with no equipartition, while $\mu = 10$ is effectively consistent with full equipartition, as it implies $m_{\mathrm{eq}} = 0.1 M_{\odot}$, which corresponds to approximately the lower mass limit found in GCs. Negative values of $\mu$ indicate that the velocity dispersion increases with stellar mass (see e.g. \cite{pavlik2022}, \cite{aros2023}, \cite{pavlik2024} and \cite{livernois2024} for simulations where clusters can show this behavior).

We employ the same fitting procedure as described in \cite{aros2023}, which follows the approach described by \cite{watkins2022} and their likelihood function. Considering a two-dimensional velocity dispersion $\sigma$, the likelihood function is defined as

\begin{equation}
\small
\begin{split}
L(\mu_R,\mu_T,m\mid\Theta) = & \prod_{i}^{N}(2\pi(\sigma(m_i\mid\Theta)^2+\delta_{\mathrm{\mu_R},i}^2))^{-1/2} \\
& \times (2\pi(\sigma(m_i\mid\Theta)^2+\delta_{\mathrm{\mu_T},i}^2))^{-1/2} \\
& \times\exp\Bigg(-\frac{1}{2}\frac{{\mathrm{\mu_R},i}^2}{\sigma(m_i\mid\Theta)^2+\delta_{\mathrm{\mu_R},i}^2} -\frac{1}{2}\frac{{\mathrm{\mu_T},i}^2}{\sigma(m_i\mid\Theta)^2+\delta_{\mathrm{\mu_T},i}^2}\Bigg)
\end{split}
\end{equation}
where $m$ is the stellar mass, $\mu_R$ and $\mu_T$ are the proper motions in the radial and tangential directions, with their associated errors $\delta_{\mu_R}$ and $\delta_{\mu_T}$, respectively. $\sigma(m_i\mid\Theta)$ is the mass-dependent velocity dispersion, with $\Theta$ representing the parameters of the chosen energy equipartition model (equations \ref{eq:classic} and \ref{eq:bianchini}).

\begin{figure*}[h]
    \centering
    \includegraphics[width=0.245\textwidth]{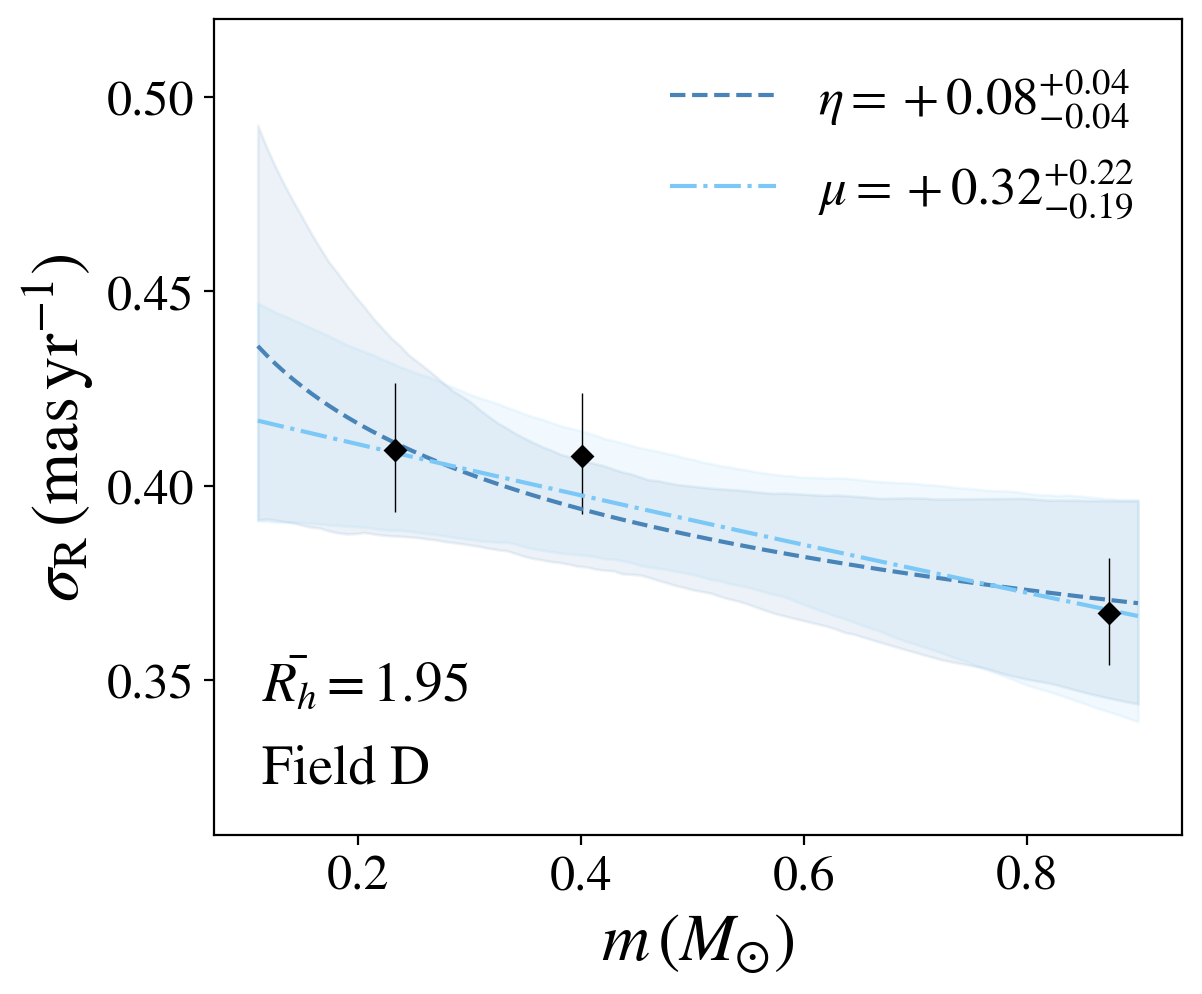}
    \hfill
    \includegraphics[width=0.245\textwidth]{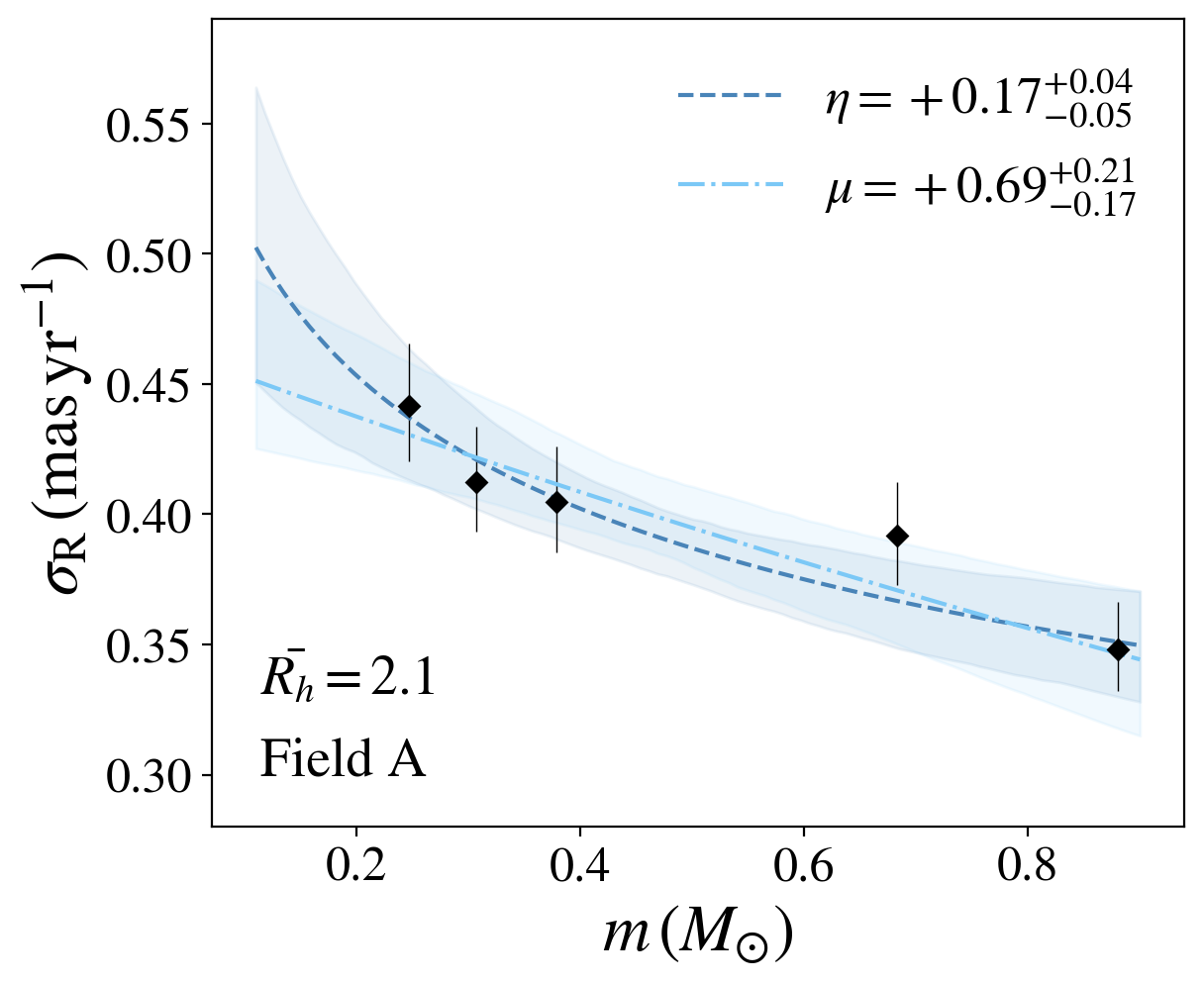}
    \hfill
    \includegraphics[width=0.245\textwidth]{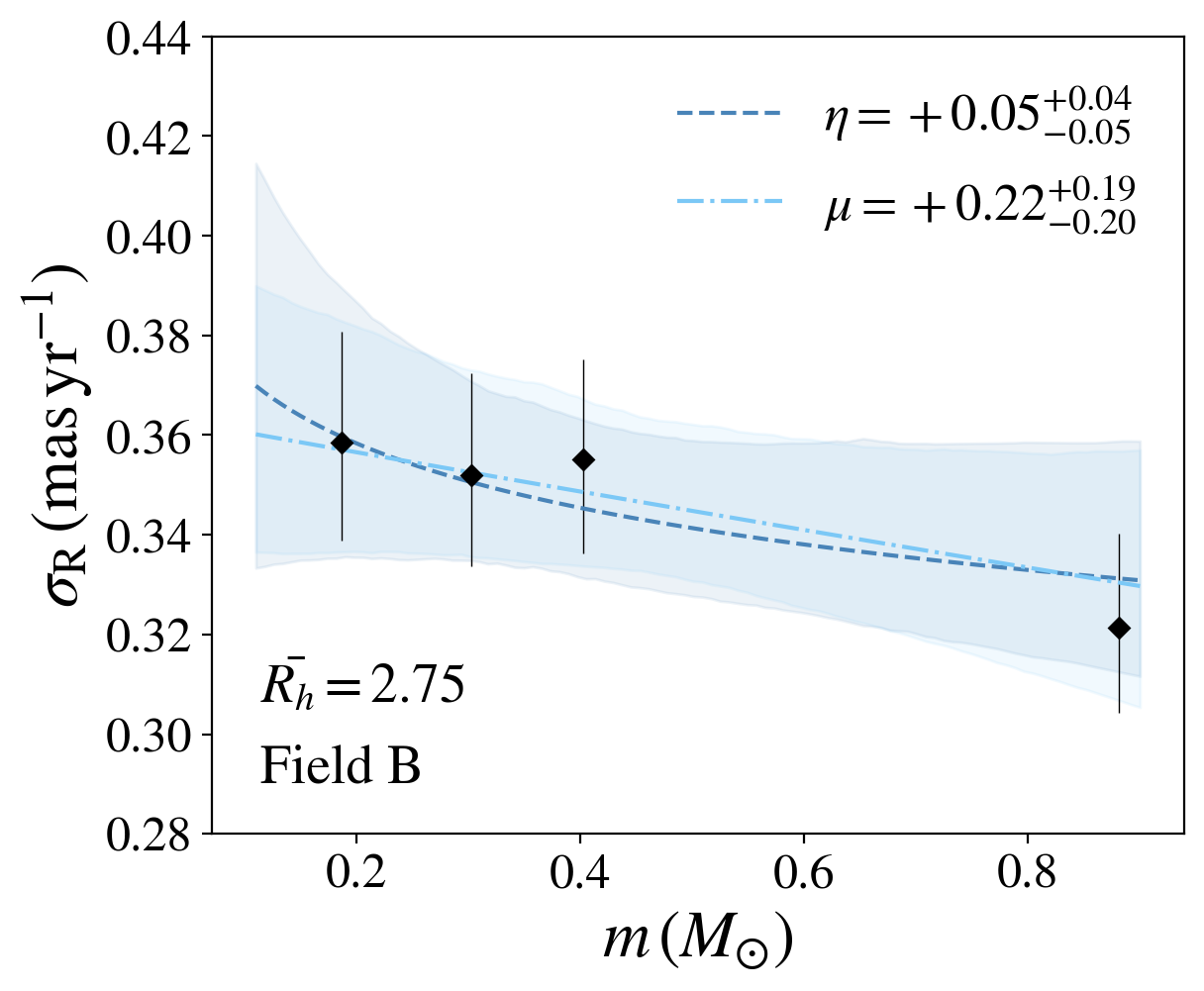}
    \hfill
    \includegraphics[width=0.245\textwidth]{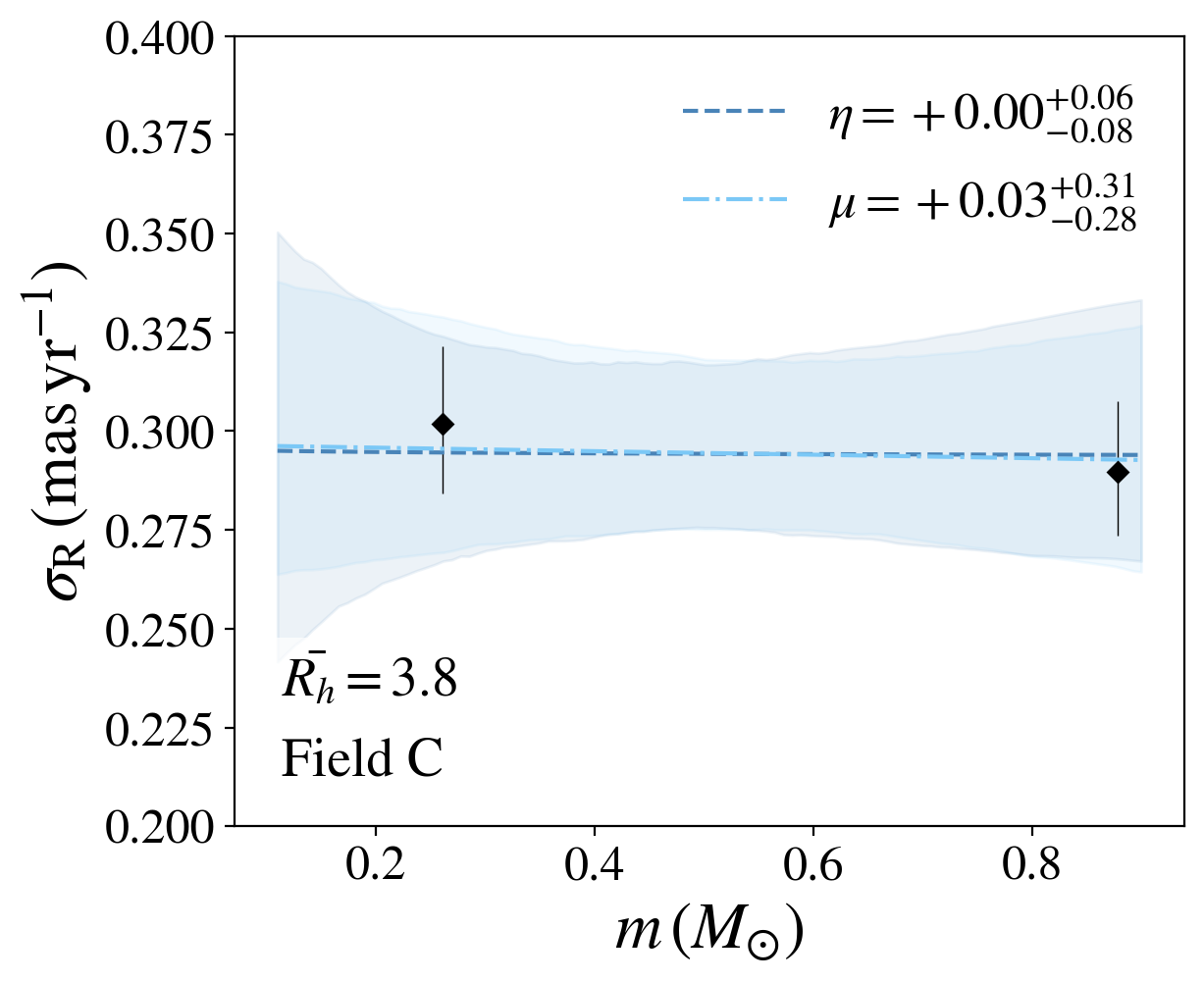}
    
    \includegraphics[width=0.245\textwidth]{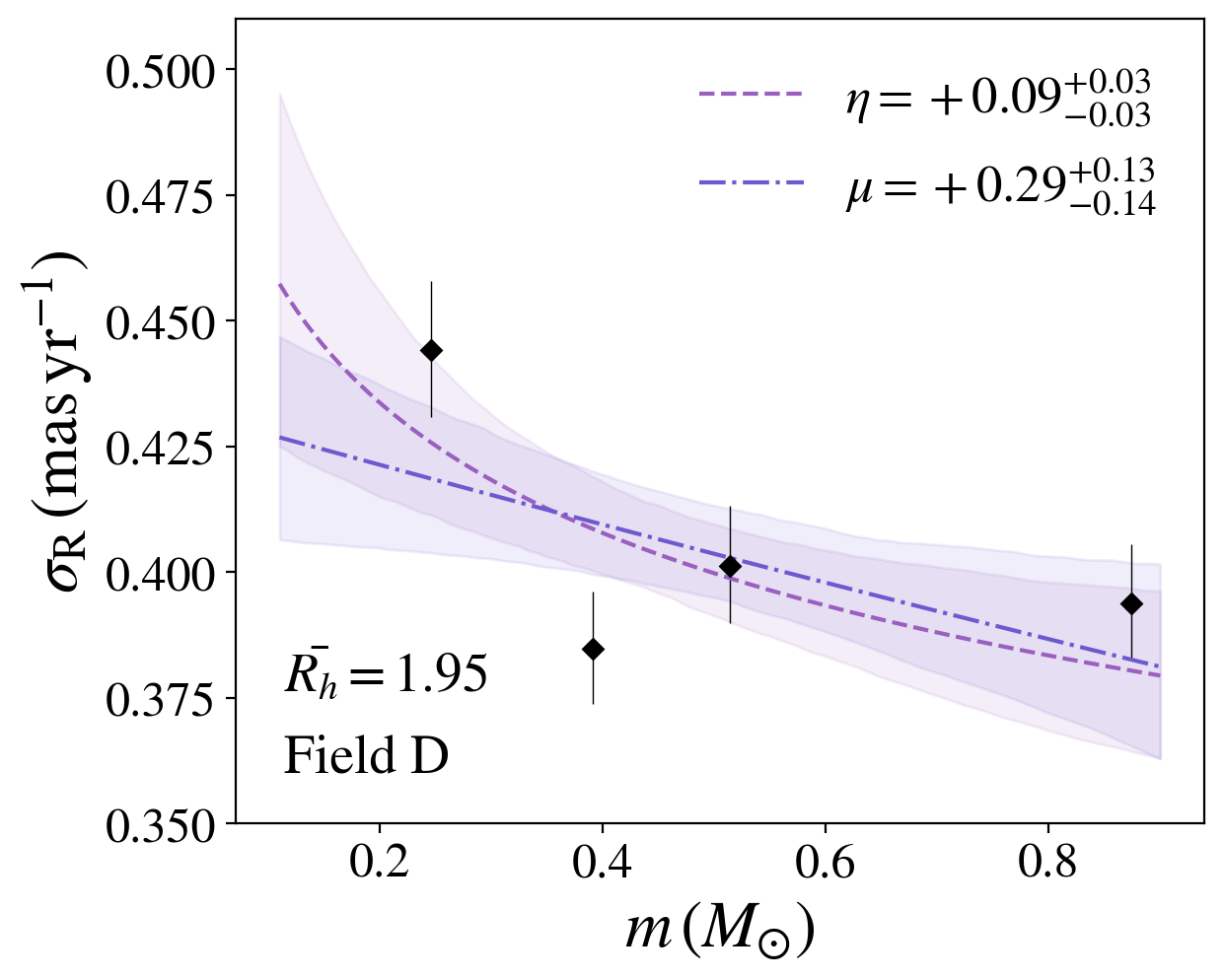}
    \hfill
    \includegraphics[width=0.245\textwidth]{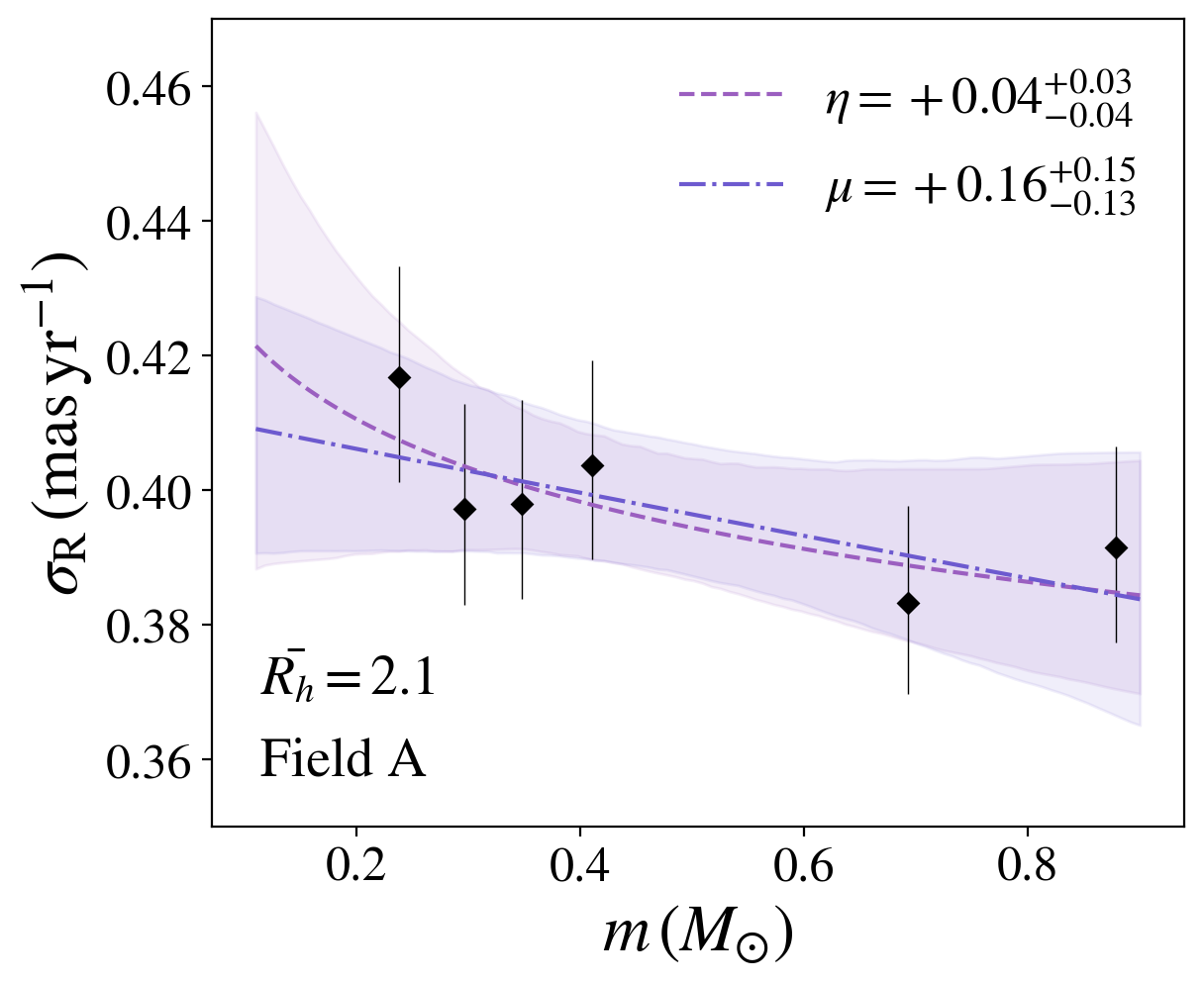}
    \hfill
    \includegraphics[width=0.245\textwidth]{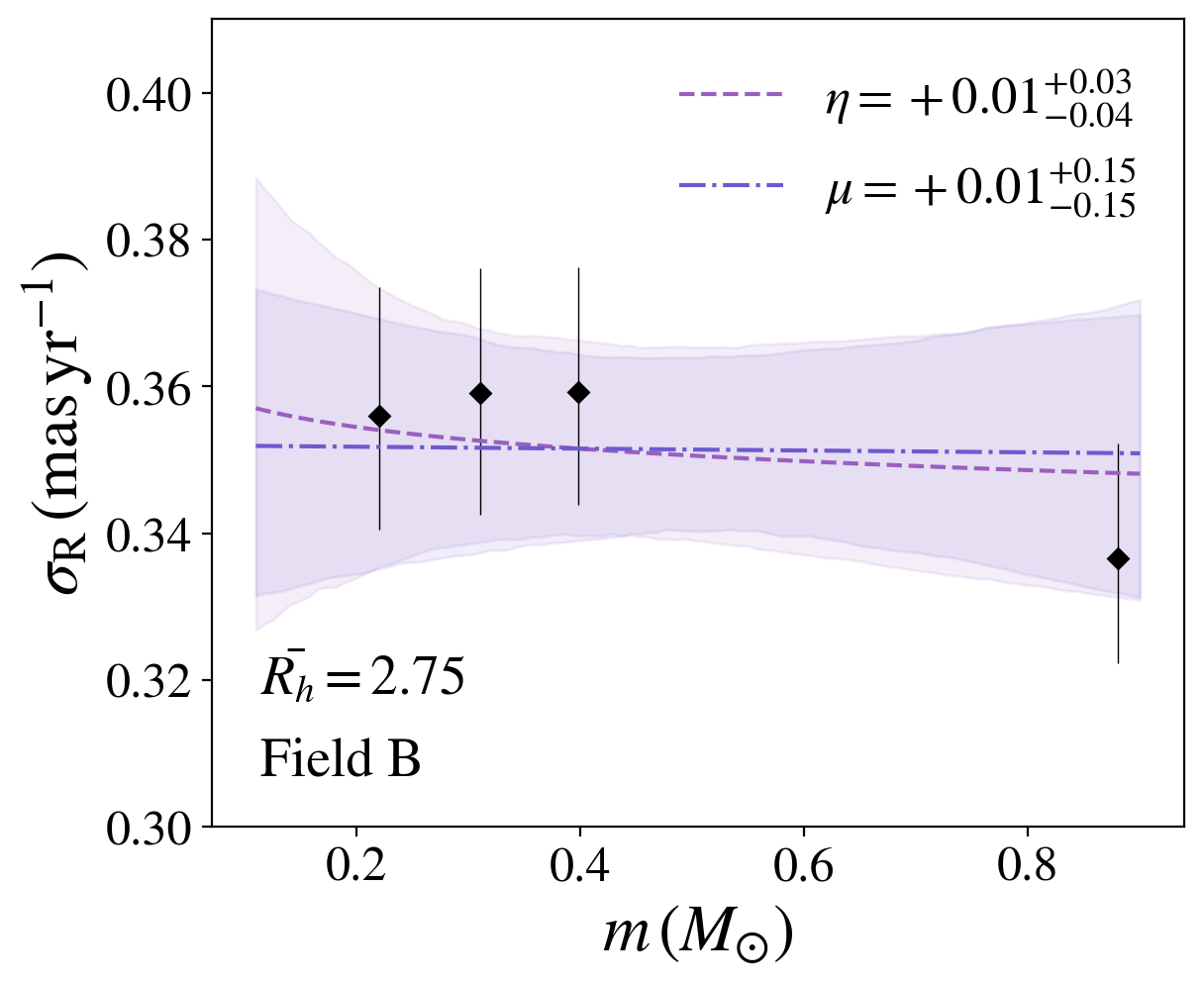}
    \hfill
    \includegraphics[width=0.245\textwidth]{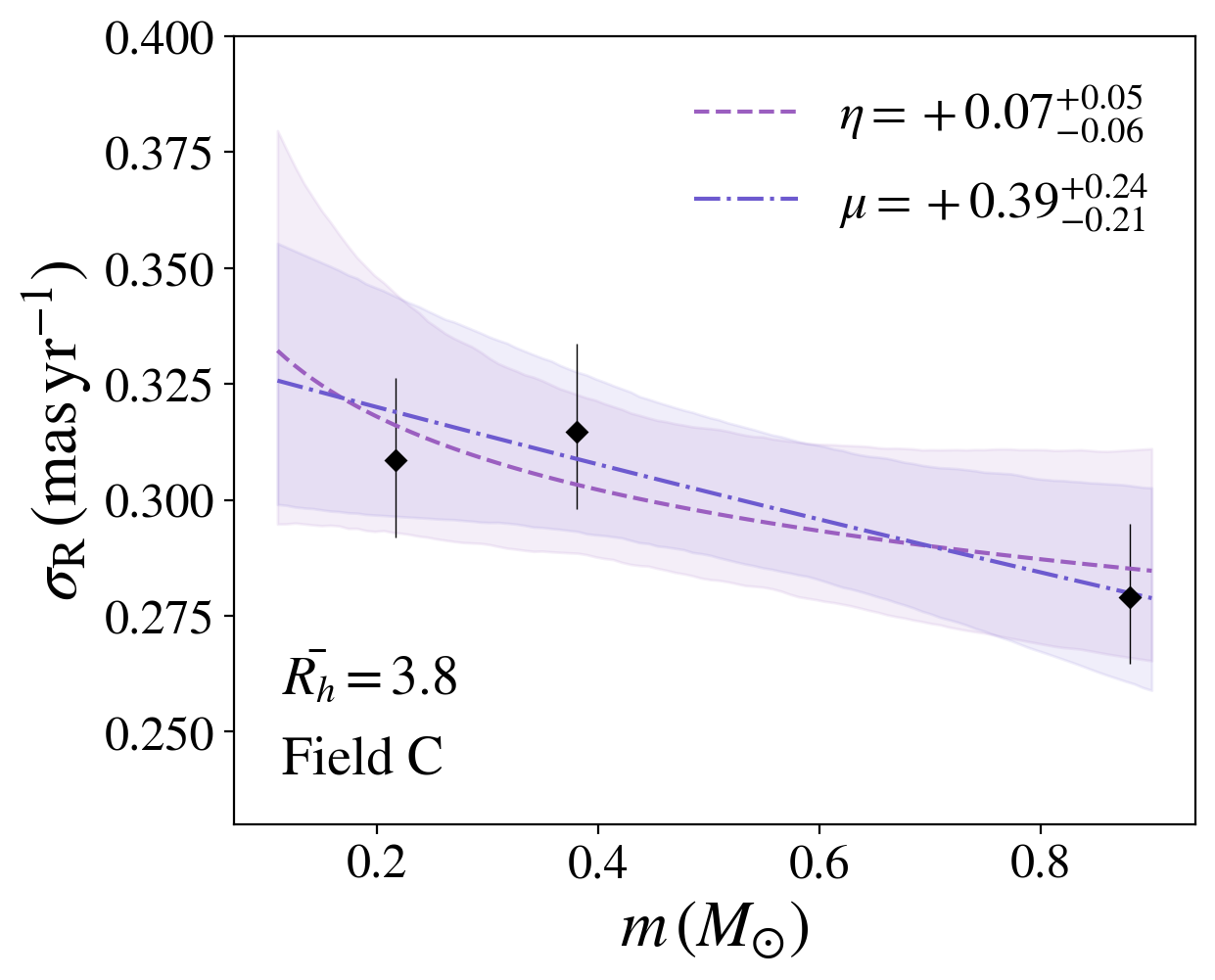}    
    \caption{Same as Figure \ref{fig:eqall}, but only considering the radial component of the velocity dispersion.}
    \label{fig:eqrad}
\end{figure*}

Figure \ref{fig:eqall} displays the total velocity dispersion as a function of stellar mass for fields A, B, C, and D combined with Gaia XP and Stetson data. The top row shows the fits of the degrees of energy equipartition $\eta$ and $\mu$ for 1G stars, while the second row shows the same for 2G stars. The uncertainties are derived from the 16th and 84th percentiles of the sampled parameter distributions, representing the lower and upper bounds of a 68\% credible interval. We present the results separately for each field, as they have different radial coverage, and energy equipartition can vary with radius, exhibiting higher values in the center of the cluster \citep{bianchini2018a}. 
The average radius for each field is indicated within the subplots. Each bin contains more than 100 stars, with denser regions containing approximately 400 stars per bin. We note that the bins are displayed for visual comparison only and not for fitting purposes.

When considering the total velocity dispersion, 1G and 2G stars show similar levels of energy equipartition in the fields D, A and B, accounting for observational errors. In the outermost field, 1G stars do not appear to be in equipartition, while 2G stars exhibit a small but significant degree of energy equipartition ($\eta = 0.09_{-0.04}^{+0.03}$ and $\mu = 0.40 \pm 0.15$). \cite{watkins2022} measured the degree of energy equipartition for all stars in the center of 47,Tucanae, finding a value of $\eta = 0.220_{-0.024}^{+0.027}$. This value is significantly higher than those obtained for the outskirts in this study, a behavior consistent with theoretical models \citep[see, e.g.,][]{livernois2024}.

Figures \ref{fig:eqtan} and \ref{fig:eqrad} are similar to Figure \ref{fig:eqall}, but for the tangential and radial components of the velocity dispersion, respectively. The evolution toward energy equipartition between 1G and 2G stars shows the largest differences in the tangential component. 
The degree of energy equipartition in the tangential component of
1G stars is consistent with no mass-velocity dispersion trend or even a slightly inverted trend with the velocity dispersion increasing with the stellar mass (see Field A, for example), while 2G stars show some degree of energy equipartition in all fields. For the radial component of the velocity dispersion, 1G and 2G stars show similar trends, except in Field A, which has a higher $\eta$ for 1G stars than 2G stars, and in the outer field, where 2G stars are instead closer to equipartition than 1G stars.

\begin{figure*}
    \centering
    \includegraphics[width=18cm]{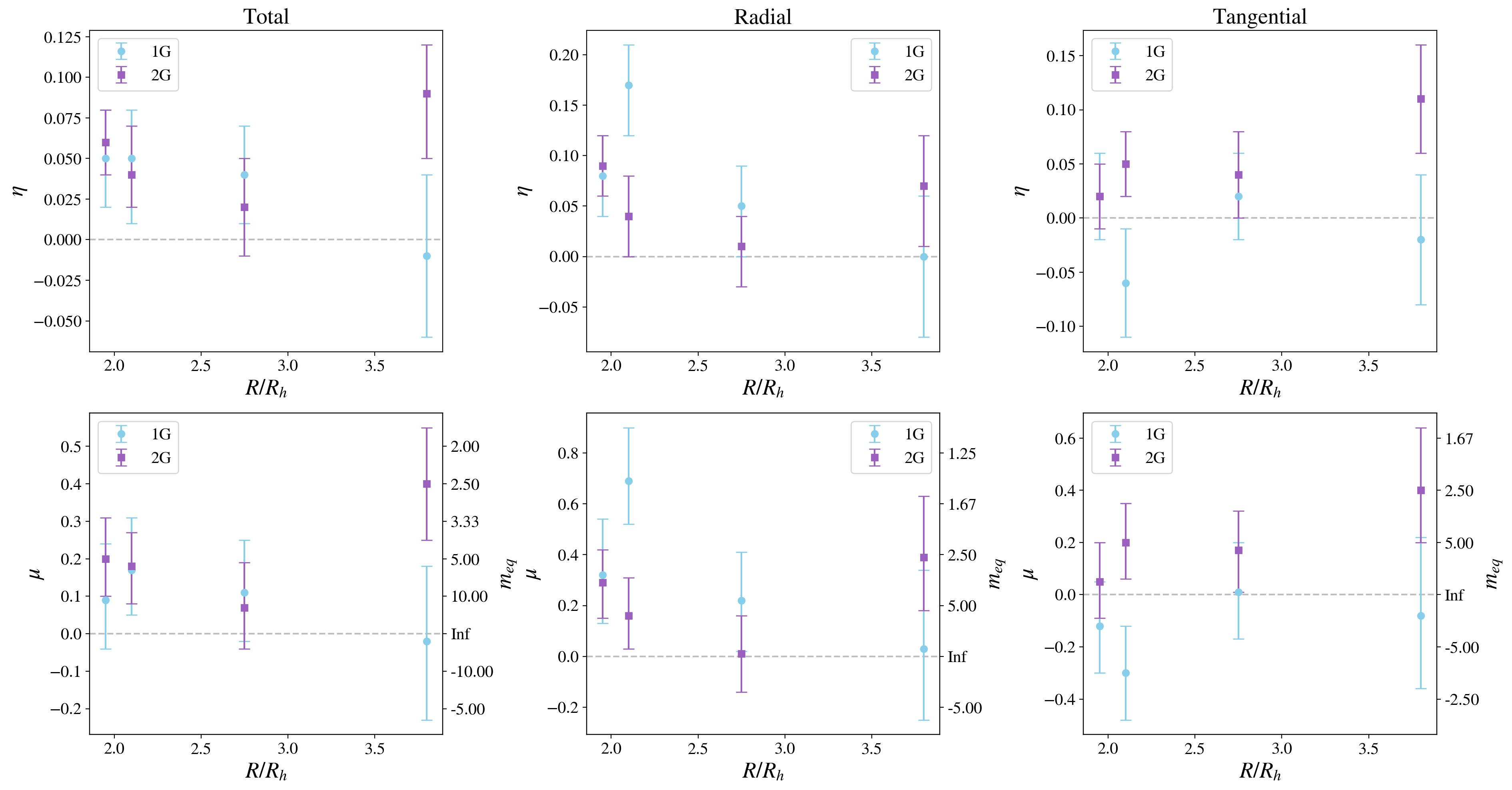}
    \caption{Best fit $\eta$ (top) and $\mu$ and $m_{eq}$ (bottom) as functions of radial distance from the cluster for 1G (blue) and 2G stars (purple). The total, radial, and tangential components are shown from left to right.}
    \label{fig:eqprofile}
\end{figure*}

Figure \ref{fig:eqprofile} 
illustrates the variation of the degree of energy
equipartition as a function of radial distance. We display, as a function of radius, $\eta$ (top row) and $\mu$ (bottom row) for the total (left panels), radial (center panels), and tangential (right panels) components of the velocity dispersion. While the degree of energy equipartition considering the total velocity dispersion is similar for 1G and 2G stars apart from the outermost field, the tangential component presents consistent differences in the evolution toward equipartition across all studied radii. While 1G stars are either consistent with not being in equipartition or exhibiting inverted equipartition, 
2G stars show a higher degree of energy equipartition. 
In the radial component, the differences between 1G and 2G stars vary with the field, with 1G stars being closer to equipartition in some fields, and the 2G stars characterized by a stronger degree of equipartition in others.

\section{Summary and discussion} \label{sec:sd}

We investigated the internal kinematics of multiple populations in 47\,Tucanae. By combining multi-epoch observations of \textit{HST} and \textit{JWST}, we derived proper motions for stars down to the H-burning limit. We combined this dataset with ground-based photometry and Gaia proper motions to characterize multiple populations in the high-mass regime. Our final dataset, which spans a wide radial and mass range, enabled us to measure energy equipartition for the multiple stellar populations of 47\,Tucanae for the first time. Moreover, we investigated the velocity dispersion and anisotropy profiles of $1G$, $2G_A$, $2G_B$, and white dwarf stars in the cluster. Our results can be summarized as follows.

\begin{itemize}
    \item 
    The observed degree of energy equipartition differs for 1G and 2G stars  and is strongly dependent on the velocity dispersion component and radial distance.
    While the total velocity dispersion shows similar equipartition trends for 1G and 2G stars apart from the outermost regions, the tangential velocity component demonstrates consistent differences across all studied radii. 
    2G stars consistently exhibit a higher degree of energy equipartition, whereas 1G stars display either no equipartition or even inverted equipartition.
    The radial velocity component shows more complex dynamics, with the equipartition progression varying across different stellar fields and generations.
    These results are consistent with the recent simulations of \cite{livernois2024} following the dynamical evolution of multiple populations with 2G stars initially more centrally concentrated than 1G stars; in
    all their simulations, \cite{livernois2024} found that the degree of energy equipartition of 2G stars is higher in the tangential component than that of 1G stars at the radial distances in this study.
    In the radial component, they also found a more variable behavior, with the differences between 1G and 2G being more dependent on the initial conditions of the simulations (see their Figure 7).

    \item 1G and 2G stars exhibit similar radial velocity dispersions along the observed radial range. However, they show significant differences in their tangential velocity dispersions. These differences translate into distinct anisotropy profiles for 1G and 2G stars. While 1G stars are isotropic, 2G stars are isotropic in the center but become radially anisotropic, with a maximum in anisotropy near 3$R_h$. They then reach isotropy again at around 10$R_h$.
    Simulations of the dynamics of multiple population that model 2G stars with an initial higher central concentration than 1G stars exhibit the same qualitative behavior for these profiles \citep[see, e.g.,][]{vesperini2021,aros2025}.

    \item We compared the radial profiles of mean $\mu_T$ for 1G and 2G stars by matching Gaia DR3 with the photometric catalog of \cite{lee2022a}, which covers an area of $1^{\circ} \times 1^{\circ}$. We did not find significant differences in these profiles. 
    However, our analysis shows that 2G stars have larger values of the ratio of rotational velocity to velocity dispersion indicating differences in the relative strength of rotational to the random motion components.
   
    \item We measured the skewness of the proper motion components and found that 1G stars exhibit higher skewness in $\mu_\mathrm{T}$ compared to 2G stars. This result provides further evidence of kinematic differences between the 1G and 2G populations and adds a new constraint for dynamical studies of multiple populations.
    We briefly discussed various dynamical processes which might affect the skewness of the velocity distribution but further investigations and dedicated numerical simulations are needed to interpret our observational results and understand the dynamical processes responsible for this difference.

    \item We analyzed the dynamical profiles of the $2G_A$ stars and the more chemically extreme $2G_B$ stars. Interestingly, while these profiles differ significantly from those of 1G stars, they are consistent with each other. Although there are indications that $2G_B$ stars may be slightly more radially anisotropic than $2G_A$ stars, the uncertainties in the data prevent us from drawing strong conclusions.

    \item Deep multi-epoch \textit{HST} observations allowed us to determine proper motions of white dwarf stars in field A. Since multiple populations could not be identified among white dwarfs, we compared the kinematics of these stars with the overall dynamical profiles of all stars in other evolutionary stages in the cluster. The large uncertainties associated with white dwarfs' proper motions prevented us from concluding if they are moving as low- or high-mass stars.
\end{itemize}

\begin{acknowledgements}
      T. Z. acknowledges funding from the European Union’s Horizon 2020 research and innovation programme under the Marie Sklodowska-Curie Grant Agreement No. 101034319 and from the European Union – NextGenerationEU. This work has been funded by the European Union – NextGenerationEU RRF M4C2 1.1 (PRIN 2022 2022MMEB9W: `Understanding the formation of globular clusters with their multiple stellar generations', CUP C53D23001200006). EV acknowledges support from NSF grant AST-2009193.
\end{acknowledgements}

\bibliography{references}
\bibliographystyle{aa}

\end{document}